\documentclass[11pt, letterpaper]{article}   	
\pdfoutput=1 
\usepackage{jheppub}
\usepackage{journals}

\usepackage{amsmath, amsthm, graphicx,amssymb,subfigure,bbm}
\usepackage[all]{xy}
\let\OLDthebibliography\thebibliography
\renewcommand\thebibliography[1]{
  \OLDthebibliography{#1}
  \setlength{\parskip}{2.5pt}
  \setlength{\itemsep}{0pt plus 0.3ex}
}

\author[]{Veselin G. Filev,}
\author[]{Denjoe O'Connor,}
\affiliation[]{  School of Theoretical Physics,\\ 
       Dublin Institute for Advanced Studies, \\
       10 Burlington Road, 
       Dublin 4, Ireland.}
\emailAdd{vfilev@stp.dias.ie}
\emailAdd{denjoe@stp.dias.ie}

\abstract{We perform computer simulations of the Berkooz--Douglas (BD)
 matrix model, holographically dual to the D0/D4--brane
 intersection. We generate the fundamental condensate versus bare
 mass curve of the theory both holographically and from simulations
 of the BD model. Our studies show excellent agreement of the two
 approaches in the deconfined phase of the theory and significant
 deviations in the confined phase. We argue the discrepancy in the
 confined phase is explained by the embedding of the D4--brane
 which yields stronger $\alpha'$ corrections to the condensate in this
 phase.}

\title{A Computer Test of Holographic Flavour Dynamics}

\begin{document}
\preprint{DIAS-STP-14-12}
\maketitle

\section{Introduction}

Gauge/gravity duality is among the most important theoretical
developments coming from string theory. In the original formulation of
Maldacena~\cite{Maldacena:1997re}, the duality relates string theory
in the $AdS_5\times S^5$ background space-time to the large $N$
limit of $3+1$ dimensional ${\cal N}=4$ Supersymmetric Yang-Mill
theory living on the asymptotic boundary of the AdS$_5$
space-time. This idea has inspired numerous extensions of the duality
with ever increasing phenomenological relevance, currently ranging
from heavy ion collisions to condensed matter physics. In this paper
we are interested in holographic flavour dynamics--the generalisation
of the AdS/CFT correspondence to flavoured gauge theories.

The first such generalisation was proposed by Karch and
Katz~\cite{Karch:2002sh} , who introduced a probe D7--brane to the
$AdS_5\times S^5$ supergravity background. On the field theory side
this corresponds to introducing an ${\cal N}=2$ fundamental
hypermultiplet in the quenched approximation. The classical dynamics
of the probe brane is governed by an effective Dirac-Born-Infeld
action. Remarkably the AdS/CFT dictionary relates the classical
properties of the brane to quantum vacuum expectation values in the
dual flavoured gauge theory. One such quantity is the fundamental
condensate of the theory, which is encoded in the classical profile of
the probe brane near the asymptotic boundary. In
refs.~\cite{Babington:2003vm}, \cite{Hoyos:2006gb} the finite
temperature set-up has been considered. The authors uncovered a first
order meson melting phase transition corresponding to a topology
change transition of the possible D7--brane embeddings. In ref.~\cite{Mateos:2007vn} 
these studies have been extended to the general
Dp/Dq--brane system and certain universal properties of the
corresponding holographic gauge theories have been uncovered.

By turning on the gauge field on the probe D--brane numerous other
control parameters can be introduced. Examples include: chemical
potential \cite{Mateos:2007vc}, external electric and magnetic fields \cite{EM-field}, isospin chemical
potential  \cite{Erdmenger:2007ja} and $R$-charge chemical potential \cite{Evans:2008nf, O'Bannon:2008bz}. This has lead to
remarkable phenomenological applications of the AdS/CFT
correspondence. However, almost exclusively these applications require
broken supersymmetry (a poorly tested regime of the duality) making
the nature of these studies somewhat speculative.\footnote{For a precision test of Gauge/Gravity duality with flavour in a supersymmetric setting see ref.~\cite{Karch:2015kfa}.} Our objective in
this paper is to perform a highly non-trivial precision test of the
gauge/gravity duality with flavours.

Testing the AdS/CFT correspondence requires an alternative
nonperturbative approach and for a four dimensional gauge theory
lattice simulations on a computer seem a natural
approach. Unfortunately, although the subject of active research \cite{Catterall:2005fd, Kaplan:2005ta}, the
lattice formulation of four dimensional Supersymmetric Yang-Mills
theory is still in its infancy. When faced with such difficulties, a
useful approach is to study simplified versions of the
correspondence. Recently progress in this direction has been made by
studying a $0+1$ dimensional version of the correspondence, one which
relates the maximally supersymmetric BFSS matrix model and its dual
type IIA supergravity background \cite{Anagnostopoulos:2007fw, Catterall:2008yz,
  Hanada:2008ez, Catterall:2009xn, Hanada:2013rga, Kadoh:2015mka, Filev:2015hia}\footnote{For a recent review we refer the reader to ref.~\cite{Joseph:2015xwa}.}. To add flavours to
this set-up we introduce a probe D4--brane. The resulting
supersymmetric quantum mechanics is knows as the Berkooz--Douglas (BD)
matrix model. Simulating the BD matrix model is one of the main
results of our paper.

Another appealing feature of the BD matrix model is that it is dual to
the D0/D4--brane system, which falls into the same universality class
\cite{Mateos:2007vn} as the phenomenologically relevant D3/D7--brane
system.

In section 2 of the paper we review the D0/D4--brane holographic
set-up. We discuss the properties of a flavour D4--brane probing the
near horizon limit of a finite temperature D0--brane supergravity
background. The model features a first order confinement/deconfinement
phase transition of the fundamental matter, which corresponds to a
topology change transition of the D4-brane embedding. This transition
can be seen as a discontinuity in the fundamental condensate as a
function of the bare mass parameter. Using the AdS/CFT dictionary
\cite{Mateos:2007vn}, we construct numerically the condensate
curve. Comparing this curve to lattice simulation is our main strategy
for testing the gauge/gravity duality.

Section 3 of the paper outlines the properties of the BFSS matrix
model and its flavoured version the BD matrix model. We describe the
Wick rotation of the DB model and the lattice discretisation that we
employ, which avoids fermion doubling. We also describe the Rational
Hybrid Monte Carlo approach to this model. A reader who is not
interested in the details of the Monte Carlo simulation can skip most
of this section and move on to section 4.

In section 4 we compare the predictions for the fundamental condensate
from both approaches: holographic studies and Monte Carlo
simulations. We perform studies at two different temperatures. Our
studies show excellent agreement between the two approaches at small
bare mass parameter. For the lower temperature this agreement extends
to the whole range of bare masses in the deconfined (black hole) phase
of the theory. We explain this by arguing that the $\alpha'$
corrections to the free energy experienced by black hole embeddings
vary weakly with the bare mass parameter and as a result largely
cancels in the calculation of the fundamental condensate, which is a
derivative of the free energy with respect to the bare mass. In the
Minkowski phase of the theory the lattice simulations deviate from the
theoretical curve for both temperatures. We argue that this reflects
the fact that $\alpha'$ corrections vary significantly with the bare
mass in this phase and hence contribute to the condensate.  The
essential difference in the two phases is that in the black hole phase
the D4--brane is restricted to pass through the black hole horizon
whereas in the Minkowski phase the embedding closes at a higher radius
that varies with the mass.  We discuss future studies to improve the
agreement in the Minkowski phase.

Our studies provide a highly non-trivial test of the AdS/CFT
correspondence with matter. Although it is not a mathematical proof,
we believe that the remarkable agreement between theory and
simulation, which we uncovered due to the cancelation mechanism
described above, provides substantial evidence for the validity of the
holographic approach to flavour dynamics.

\section{Holographic flavours in one dimension}

In this section we focus on the D0/D4--brane system. This system is
particularly attractive for a precision test of holography since on
one side the corresponding dual gauge theory is one dimensional,
making it accessible via computer simulations and on the other side it
is in the same universality class as the D3/D7--brane system, which
plays a key role in holographic flavour dynamics. In what follows we
briefly review the description of this system in the quenched
approximation adapting the general discussion of
references~\cite{Mateos:2007vn} and~\cite{Itzhaki:1998dd}.

\subsection{D0-brane background}
In the near horizon limit the D0-brane background is given by the
metric:
\begin{eqnarray}\label{metric}
ds^2&=&-H^{-\frac{1}{2}}\,f\,dt^2+H^{\frac{1}{2}}\left(\frac{du^2}{f}+u^2\,d\Omega_8^2\right)\ ,\nonumber \\
e^{\Phi}&=&H^{\frac{3}{4}}\ ,~~~~~~C_0 =H^{-1}\ ,
\end{eqnarray}
where $H=\left(L/u\right)^7$, $f(u) = 1-(u_0/u)^{7}$ is the blackening factor, $\Phi$ is the dilaton field and $C_0$ is the only component of the RR one form coupled to the D0-branes . Here $u_0$ is the radius of the horizon and the length scale $L$ can be expressed in terms of string theory units as:
\begin{equation}
L^7=60\,\pi^3\,g_s\,N_c\,\alpha'^{7/2}\ ,
\end{equation}
where $N_c$ is the number of D0--branes corresponding to the rank of the gauge group of the dual field theory\footnote{Note that we will abbreviate $N_c$ to $N$ when the context is clear.}. According to the general gauge/gravity duality \cite{Itzhaki:1998dd}, the Yang-Mills coupling of the corresponding dual gauge theory is given by:
\begin{equation}
g_{\rm YM}^2=g_s\,(2\pi)^{-2}\,\alpha'^{-3/2}\ .
\end{equation}
The Yang-Mills coupling is dimensionful and the corresponding dimensionless effective coupling runs with the energy scale according to:
\begin{equation}\label{geff}
g_{\rm eff}^2=\lambda\,U^{-3}\ ,
\end{equation}
where $\lambda = g_{\rm YM}^2\,N_c$ is the t'Hooft coupling. The supergravity background can be trusted if both the curvature and the dilaton are small, which leads to the restriction \cite{Itzhaki:1998dd}:
\begin{equation}\label{bounds}
1\ll g_{\rm eff} \ll N_c^{\frac{4}{7}}\ .
\end{equation}
and the theory is strongly coupled in this regime. From equations (\ref{metric}) and (\ref{geff}) it follows that the upper bound in equation (\ref{bounds}) can be violated at low energies (small radial distances) when the dilaton blows, however at finite temperature and fixed 'tHooft coupling, $g_{\rm eff}$ peaks at the black hole horizon and the bound $\lambda/T^3 \ll N_c^{8/7}$ is satisfied in the large $N$ limit. At high energies (large radial distances) the curvature of the background grows, while the effective coupling decreases. As a result the lower bound in (\ref{bounds}) is violated at energies $U\gtrsim \lambda^{1/3}$ and hence $\alpha'$ corrections are increasingly important at large radial distances.

Finally, the Hawking temperature of the background is given by:
\begin{equation}
T =\frac{7}{4\,\pi\,L}\left(\frac{u_0}{L}\right)^{\frac{5}{2}}
\end{equation}
and is identified with the temperature of the dual gauge theory.

\subsection{Flavour D4-branes}
\label{holo-condensate}
To introduce matter in the fundamental representation we consider the addition of $N_f$ D4-branes to the D0-brane background. In the probe approximation $N_f\ll N_c$, the dynamics of the D4-branes is governed by the Dirac-Born-Infeld, which in the absence of a background B-field is given by:
\begin{equation}\label{DBI}
S_{\rm DBI} = -N_f\,T_4\int\,d^4\xi\,e^{-\Phi}\,\sqrt{-{\rm det}||G_{\alpha,\beta}+(2\pi\alpha')F_{\alpha,\beta}||}\ ,
\end{equation}
where $G_{\alpha,\beta}$ is the induced metric and $F_{\alpha,\beta}$ is the $U(1)$ gauge field of the D4-brane, which we will set to zero. The D4-brane tension is given by:
\begin{equation}\label{tension}
T_4=\frac{\mu_4}{g_s}=\frac{1}{(2\pi)^4\,\alpha'^{5/2}\,g_s}\ .
\end{equation}
The D4-brane embedding that we consider extends along the radial and time directions and wraps an $S^3$ sphere in the directions transverse to the D0-brane. To parametrise it let us split the unit $S^8$ in the metric (\ref{metric}) into:
\begin{equation}
d\Omega_8^2=d\theta^2+\cos^2\theta\,d\Omega_3^2+\sin^2\theta\,d\Omega_4^2\ .
\end{equation}

Our embedding now extends along $t$ and $\Omega_3$ and has a non-trivial profile in the $(u,\theta)$ plane, which we parametrise as $(u,\theta(u))$. Next we Wick rotate the action (\ref{DBI}) and periodically identify time with period $\beta =1/T$. Using equation (\ref{metric}) we obtain:
\begin{equation}\label{DBI-Wick}
S_{\rm DBI}^E =\frac{N_f\,\beta}{8\,\pi^2\,\alpha'^{5/2}\,g_s}\int\, du\,u^3\cos^3\theta(u)\,\sqrt{1+u^2\,f(u)\,\theta'(u)^2}\ .
\end{equation}
In the limit of zero temperature ($u_0\to0$) the regular solution to the equation of motion for $\theta(u)$ is given by $u\,\sin\theta = m$, where the constant $m$ is proportional to the bare mass of the flavours \cite{Karch:2002sh}, \cite{Mateos:2007vn}. At finite temperature the separation $L(u)=u\,\sin\theta(u)$ has a non-trivial profile reflecting the non-vanishing condensate of the theory. To analyse this case it is convenient to define dimensionless radial coordinate $\tilde u = u/u_0$. At large $\tilde u$ the general solution $\theta(\tilde u)$ has the expansion: 
\begin{equation}\label{Expansion-sin(theta)}
\sin\theta =\frac{\tilde m}{\tilde u}+\frac{\tilde c}{u^3}+\dots.
\end{equation}
Holography relates the dimensionless constants $\tilde m, \tilde c$ to the bare mass and condensate of the theory via \cite{Mateos:2007vn}\footnote{Note that our expressions differ slightly from the ones presented in \cite{Mateos:2007vn} due to the different choice of radial variable.}:
\begin{eqnarray}
m_q &=& \frac{u_0\,\tilde m}{2\pi\alpha'}=\left(\frac{120\,\pi^2}{49}\right)^{1/5}\left(\frac{T}{\lambda^{1/3}}\right)^{2/5}\lambda^{1/3}\,\tilde m\ , \nonumber \\
\langle {\cal O}_m\rangle &=&-\frac{N_f\,u_0^3}{2\,\pi\,g_s\,\alpha^{3/2}} \,\tilde c=\left(\frac{2^4 \,15^3\,\pi^6}{7^6}\right)^{1/5}\,N_f\,N_c\,\left(\frac{T}{\lambda^{1/3}}\right)^{6/5}\,(-2\,\tilde c)\ .
\label{dictionary}
\end{eqnarray}
We refer the reader to Appendix A for derivation of (\ref{dictionary}). Note that equation (\ref{Expansion-sin(theta)}) implies that the D7-branes are described by a one parameter family of embeddings (parametrised by $\tilde m$). In the case of the D3/D7 system this is natural due to the scaling symmetry (everything depends on the dimensionless ratio $m_q/T$), but is this consistent with the D0/D4 system, which has a dimensionful 'tHooft coupling? Indeed, the D0/D4 system has a 'tHooft coupling $\lambda$ of dimension three suggesting that there are two independent dimensionless parameters $m_q/\lambda^{1/3}$ and $T/\lambda^{1/3}$.  However, while the holographic set-up allows rescaling of the radial coordinate by $u_0$ and the description of D7-brane embeddings by the single parameter $\tilde m$, as can be seen from the first equation (\ref{dictionary}) we have $\tilde m \sim (m_q/\lambda^{1/3}) (T/\lambda^{1/3})^{-2/5}$. As a result the condensate in the second equation in (\ref{dictionary}) is indeed a function of the two dimensionless parameters $(T/\lambda^{1/3},\, m_q/\lambda^{1/3})$ which is consistent with dimensional analysis and is in contrast to the D3/D7 system, where the condensate depends on the dimensionless parameter $m_q/T$.   
\subsection{Fundamental condensate}
\begin{figure}[t] 
   \centering
   \includegraphics[width=5.5in]{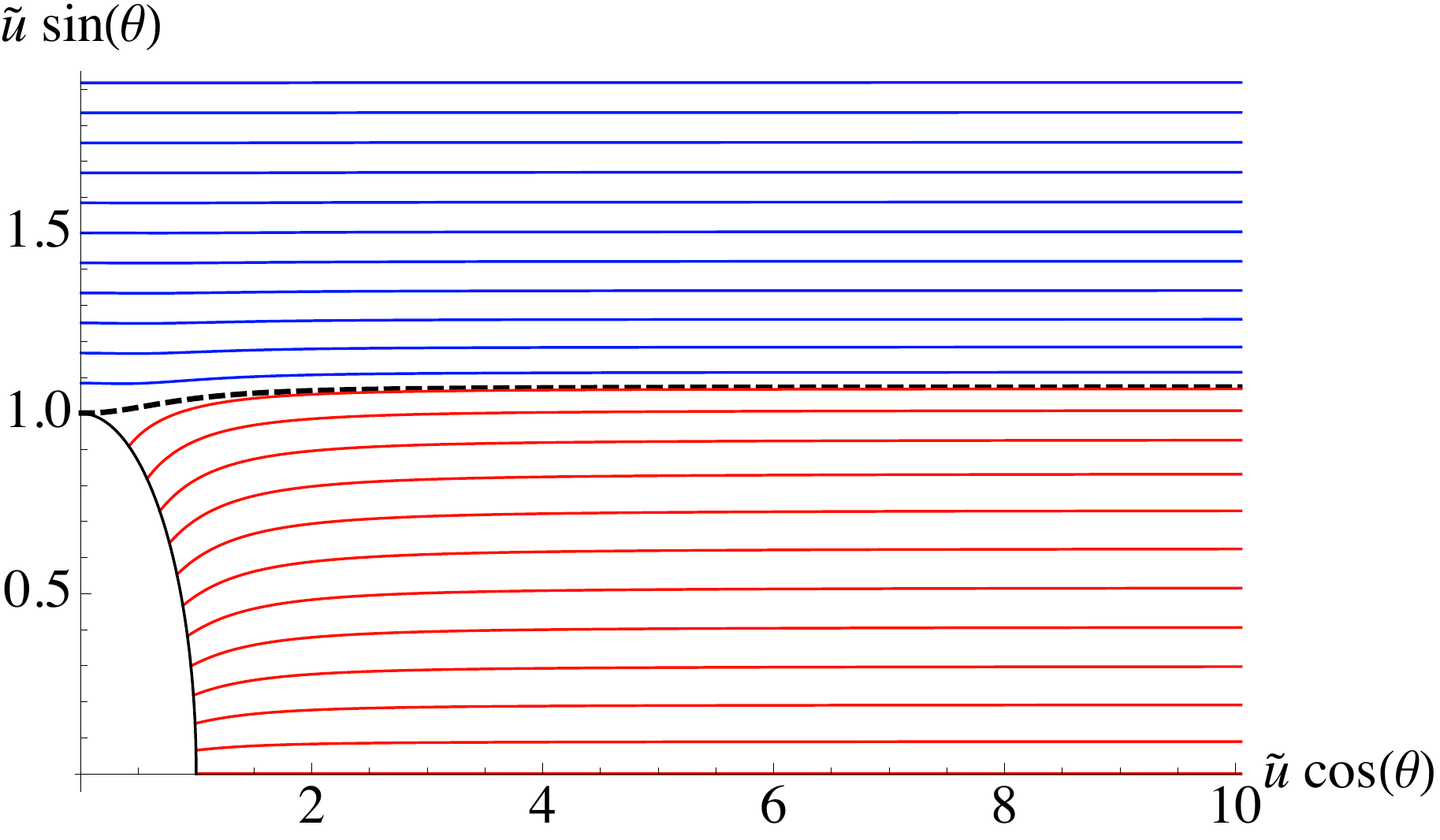} 
 \caption{\small A plot of sample Minkowski (blue curves) and black hole (red curves) embedding. The dashed curve represents the critical embedding and the black circle represents the horizon.}
   \label{fig:1}
\end{figure}

The fundamental condensate can be obtained numerically by solving the differential equation for $\theta(\tilde u)$ obtained by varying the Lagrangian:
\begin{equation}
\tilde{\cal L}\propto \tilde u^3\cos^3\theta(\tilde u)\,\sqrt{1+\tilde u^2\,(1-1/\tilde u^7)\,\theta'(\tilde u)^2}\ .
\end{equation}
The possible solutions split into two classes (look at figure \ref{fig:1}). The first class comprises of embeddings closing above the horizon at some minimal radial distance $\tilde u_{min} >1$, for such embeddings the wrapped $S^3$ sphere shrinks to zero size and hence $\theta(\tilde u_{min}) = \pi/2$. Following the terminology of ref.~\cite{Mateos:2007vn} we call these embeddings Minkowski embeddings. The spectrum of Minkowski embeddings is characterised by discrete normal modes corresponding to meson-like bound states and they are identified with the confined (bound) phase of the theory. The other class of embeddings correspond to probes which reach the horizon. They are parametrised by the size of the $S^3$ sphere at the horizon or equivalently by $\theta_0 =\theta(1)$. We refer to these embeddings as the black hole embeddings \cite{Mateos:2007vn}. Their spectrum is characterised by discrete quasi normal modes corresponding to melting mesons~\cite{Hoyos:2006gb}, and they are identified with the deconfined phase of the theory. The two classes are separated by the critical embedding satisfying $\theta_0=\pi/2$, which has a conical singularity at the horizon~\cite{Mateos:2007vn}. The topology change transition between Minkowski and black hole embeddings corresponds to a confinement/deconfinement phase transition for the fundamental matter~\cite{Babington:2003vm}. The nature of the phase transition depends on the properties of the solutions near the critical embedding. It turns out that the structure of the solutions depends only on the dimensionality of the internal $S^n$ sphere wrapped by the embedding ($n=3$ in our case). In this sense the holographic gauge theories dual to the Dp/Dq brane set-up split into universality classes (characterised by $n$)~\cite{Mateos:2007vn}. According to this nomenclature the D3/D7 system is in the same universality class as the D0/D4 system that we consider. For $n=3$ the solutions near the critical embedding have a multivalued nature and the phase transition is of a first order~\cite{Mateos:2007vn}, \cite{First-order}.

To obtain the condensate versus bare mass equation of state $\tilde c(\tilde m)$ one can read off the asymptotics of the numerical solution at large $\tilde u$ and use the holographic dictionary (\ref{dictionary}). To obtain the solutions one uses a numeric shooting technique from the bulk of the geometry. For black hole embeddings one can show that at the horizon the differential equation for $\theta(\tilde u)$ effectively reduces order and regularity completely determines the Cauchy initial conditions in terms of $\theta_0$. For Minkowski embeddings it is convenient to consider the field $\chi =\sin\theta$. Similarly to the black hole case, the differential equation for $\chi(\tilde u)$ reduces order at $\tilde u_{min}$ and the Cauchy initial conditions are completely determined by the parameter $\tilde u_{min}$. 

\begin{figure}[t] 
   \centering
   \includegraphics[width=5.5in]{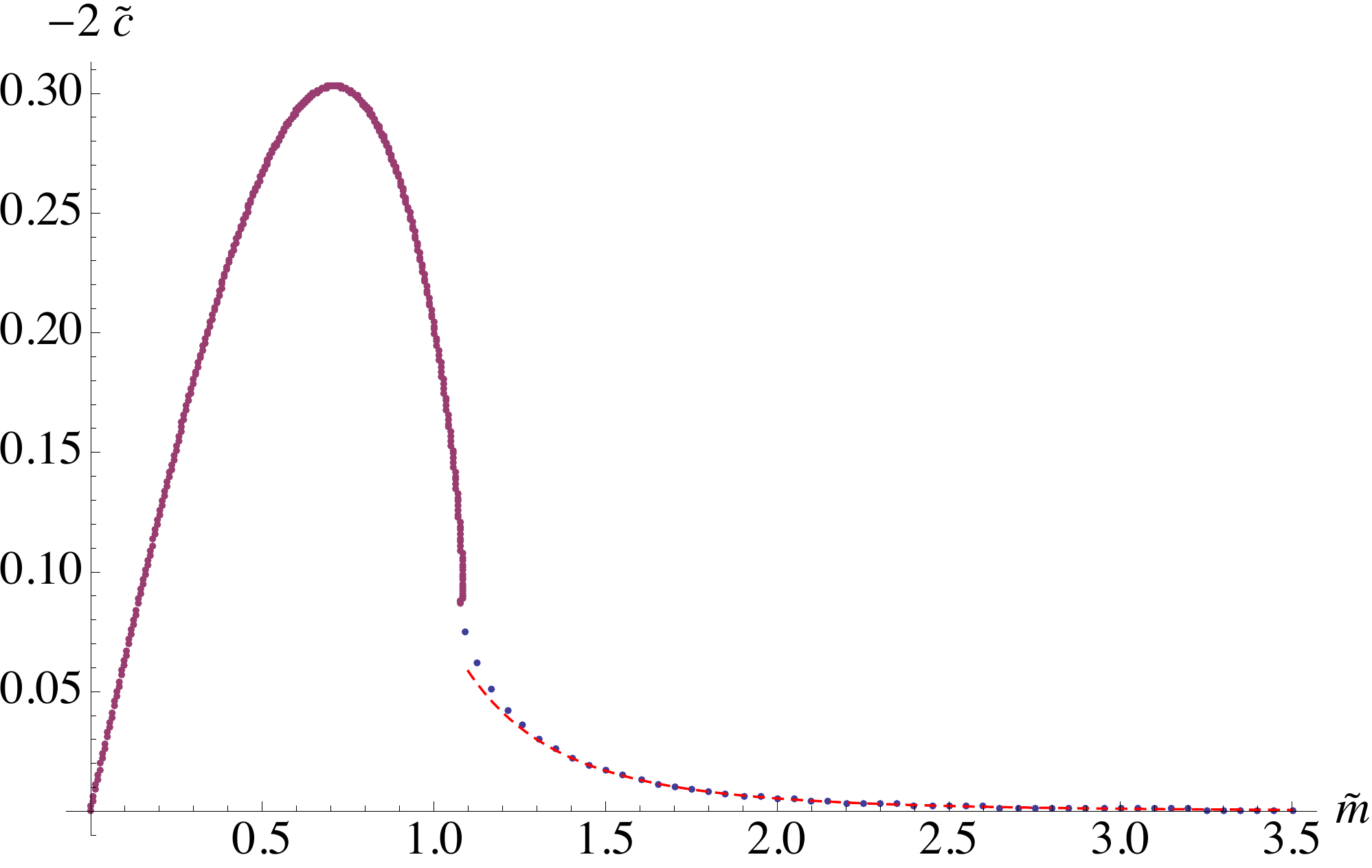} 
 \caption{\small A plot of the condensate versus bare mass curve. The red curve represents the deconfined (black hole) phase of the theory. The dotted curve corresponds to the confined (Minkowski phase) and the red dashed curve is the leading large $\tilde m$ behaviour of the condensate $\tilde c = 3/(35\,\tilde m^4)$.}
   \label{fig:2}
\end{figure}

These considerations allow for the numerical construction of the condensate versus bare mass curve $\tilde c(\tilde m)$, which we present in figure~\ref{fig:2}. The solid red curve represents the deconfined phase of the theory, corresponding to the black hole embeddings from figure \ref{fig:1}. The dotted blue curve represents the confined phase of the theory, corresponding to the Minkowski embeddings in figure \ref{fig:1}. Finally, the red dashed curve represent the analytic result $\tilde c = 3/(35\,\tilde m^4)$ for the fall-off of the condensate at large $\tilde m$. 

The goal of our study is to verify the condensate curve in figure~\ref{fig:2}, using numerical lattice simulations. Note that this curve was obtained without taking into account $\alpha'$ corrections to either the supergravity background or the DBI action. As the studies of the BFSS model performed in refs.~\cite{Hanada:2008ez}, \cite{Kadoh:2015mka} and \cite{Filev:2015hia}  show, $\alpha'$ corrections to the D0-brane background are crucial for comparing with Monte Carlo simulations. Our main observation concerning these corrections is that in the black hole phase (as can be seen from figure \ref{fig:1}) the D4-brane embeddings explore the same range of the bulk of the D0-brane background and hence the $\alpha'$ corrections to their free energy is roughly the same. This is in contrast to Minkowski embeddings whose minimal radial distance $\tilde u_{min}$ varies with the parameter $\tilde m$. Since the condensate is defined as a derivative of the free energy with respect to the bare mass parameter (see Appendix A for details), this suggests that the $\alpha'$ corrections to the condensate mostly cancel in the black hole phase and are significant in the Minkowski phase. Indeed, our numerical studies in section \ref{section test} supports this and we observe an excellent agreement in the black hole phase.

\section{Berkooz--Douglas matrix model}
The Berkooz--Douglas matrix model~\cite{Berkooz:1996is} was originally proposed as a matrix model for DLCQ M-theory with $N_c$ units of momentum along the longitudinal directions in the presence of $N_f$ longitudinal M5-branes. In this work we are interested in the interpretation of the model as the low energy effective field theory governing the D0/D4-brane system with $N_c$ (colour) D0-branes and $N_f$ (flavour) D4-branes. 

The Berkooz--Douglas matrix model represents the unique way to introduce fundamental matter to the BFSS matrix model, while preserving half of the original supersymmetry. In the following we first introduce briefly the BFSS matrix model and then discuss the introduction of fundamental degrees of freedom to the BFSS model in the six dimensional notations of ref.~\cite{VanRaamsdonk:2001cg}. Finally, we describe the lattice discretisation of the Berkooz--Douglas matrix model and its Monte Carlo simulations.

\subsection{BFSS matrix model}

The easiest way to obtain the BFSS matrix model is via dimensional reduction of ten dimensional supersymmetric Yang-Mills theory down to one dimension. We obtain:
\begin{equation}\label{BFSS 10Mink}
S_M =\frac{1}{g^2}\int dt \,{\rm tr}\left\{\frac{1}{2}({\cal D}_0X^i)^2 +\frac{1}{4}[X^i,X^j]^2-\frac{i}{2}\Psi^T C_{10}\,\Gamma^0D_0\Psi +\frac{1}{2}\Psi^T C_{10}\,\Gamma^i[X^i,\Psi]\right\}\ ,
\end{equation}
where $\Psi$ is a thirty two component Majorana--Weyl spinnor, $\Gamma^\mu$ are the 10D gamma matrices and $C_{10}$ is the charge conjugation matrix satisfying $C_{10} \Gamma^{\mu}C_{10}^{-1} = -{\Gamma^{\mu}}^T$. To avoid fermion doubling we take a representation \cite{Catterall:2008yz,Filev:2015hia} in terms of the nine dimensional gamma matrices~$\gamma^i$: 
\begin{eqnarray}\label{gammas}
\Gamma^i &=& \gamma^i\otimes \sigma_1\ , ~~~~{\rm for }~ i = 1,\dots ,9\ , \nonumber \\
\Gamma^0 &=& 1_{16}\otimes i\sigma_2\ , \nonumber \\
C_{10} &=&C_{9}\otimes i\sigma_2\ ,
\end{eqnarray}
where $C_9$ is the charge conjugation matrix in nine dimensions satisfying $C_{9} \gamma^{i}C_{9}^{-1} = {\gamma^{i}}^T$.

Next we wick rotate the action (\ref{BFSS 10Mink}) by sending: $dt \to -id\tau$, $\partial_t\to i\partial_\tau$ and $\Gamma^0\to -i\Gamma^{\tau}$. For the Euclidean action $S_E = -iS_M$ (defined to have a positive kinetic term) we obtain:
\begin{equation}
S_E =\frac{1}{g^2}\int d\tau \,{\rm tr}\left\{\frac{1}{2}({\cal D}_\tau X^i)^2 -\frac{1}{4}[X^i,X^j]^2+\frac{i}{2}\Psi^T C_{10}\,\Gamma^\tau {\cal D}_\tau\Psi -\frac{1}{2}\Psi^T C_{10}\,\Gamma^i[X^i,\Psi]\right\}\ .
\end{equation}
Note that the wick rotated fermions no longer satisfy reality condition but can still be taken to be Weyl, we thus consider the ansatz:
\begin{equation}\label{Weyl}
\Psi = \psi \otimes \left(\begin{array}{c} 1 \\ 0 \end{array}\right)\ .
\end{equation}
Using this anstatz together with equation (\ref{gammas}) and the fact that $\Gamma^{\tau} = i\Gamma^0$ for the euclidean action in 9D notation we obtain:
\begin{equation}\label{BFSS-9D}
S_E =\frac{1}{g^2}\int d\tau \,{\rm tr}\left\{\frac{1}{2}(D_\tau X^i)^2 -\frac{1}{4}[X^i,X^j]^2+\frac{1}{2}\psi^T C_{9}\,D_\tau\psi -\frac{1}{2}\psi^T C_{9}\,\gamma^i[X^i,\psi]\right\}\ ,
\end{equation}
 as one can see in this notation the action is explicitly SO(9) invariant. Note that we have not imposed any restriction on the nine  dimensional basis. For example if we choose $\gamma^i$ to be in the Majorana representation  this would imply that $C_9=1_{16}$, which is the most popular formulation of the model. However, we are interested in a basis in which the discrete theory is free of fermion doubling and one can show \cite{Catterall:2008yz} that if $C_9 = 1_8\otimes \sigma_1$ the discrete theory is indeed free of doublers. Constructing a basis for which $C_9$ is of this form is relatively straightforward. For example one can tensor up the Mayorana basis in seven dimensions $\tilde\gamma_E^a$:
\begin{eqnarray}\label{9gammas}
\gamma^a &=& -\tilde \gamma_E^a \otimes \sigma_3\ , ~~~{\rm for}~a=1,\dots,7\ ,\nonumber \\
\gamma^8 &=& 1_8\otimes\sigma_2\ ,\nonumber \\
\gamma^9 &=& 1_8\otimes\sigma_1\ ,
\end{eqnarray}
and verify that indeed $C_9$ is of the desired form (it also satisfies $C_9=\gamma^9$). The discretisation of the action (\ref{BFSS-9D}) was considered in ref.~\cite{Filev:2015hia}, for completeness we provide the details in appendix B.

\subsection{Adding flavours}

In this section we will consider the addition of flavours to the BFSS matrix model. The resulting matrix model is know as the Berkooz--Douglas matrix model. We will follow closely the notation of ref.~\cite{VanRaamsdonk:2001cg}. This requires a basis for the ten dimensional clifford algebra, which is better suited for reduction to six dimensions. Let us begin by writing the Lagrangian of the action:
\begin{eqnarray}\label{Berkooz--Douglas}
{\cal L} &=& \frac{1}{g^2}{\rm Tr}\left(\frac{1}{2}D_0X^aD_0X^a+\frac{i}{2}\lambda^{\dagger\,\rho}D_0\lambda_\rho+\frac{1}{2}D_0{\bar X}^{\rho\dot\rho}D_0X_{\rho\dot\rho}+\frac{i}{2}\theta^{\dagger\dot\rho}D_0\theta_{\dot\rho}\right)\nonumber \\
&&+\frac{1}{g^2}{\rm tr}\left(D_0\bar\Phi^{\rho}D_0\Phi_\rho+i\chi^{\dagger}D_0\chi\right)+{\cal L}_{\rm int}\ ,
\end{eqnarray}
where:
\begin{eqnarray}\label{Berkooz--Douglas_int}
{\cal L}_{\rm int} &=&\frac{1}{g^2}{\rm Tr}\left(\frac{1}{4}[X^a,X^b][X^a,X^b]+\frac{1}{2}[X^a,\bar X^{\rho\dot\rho}][X^a,X_{\rho\dot\rho}] -\frac{1}{4}[\bar X^{\alpha\dot\alpha},X_{\beta\dot\alpha}][\bar X^{\beta\dot\beta},X_{\alpha\dot\beta}]\right) \nonumber \\
&&-\frac{1}{g^2}{\rm tr}\left(\bar\Phi^{\rho}(X^a-m^a)(X^a-m^a)\Phi_\rho\right) \nonumber \\
&&+\frac{1}{g^2}{\rm tr}\left(\bar\Phi^{\alpha}[\bar X^{\beta\dot\alpha},X_{\alpha\dot\alpha}]\Phi_\beta+\frac{1}{2}\bar\Phi^\alpha\Phi_\beta \bar\Phi^\beta\Phi_\alpha-\bar\Phi^\alpha\Phi_\alpha \bar\Phi^\beta\Phi_\beta\right) \nonumber \\
&&+\frac{1}{g^2}{\rm Tr}\left(\frac{1}{2}\bar\lambda^\rho\gamma^a[X^a,\lambda_\rho]+\frac{1}{2}\bar\theta^{\dot\alpha}\gamma^a[X^a,\theta_{\dot\alpha}]-\sqrt{2}\,i\,\varepsilon_{\alpha\beta}\,\bar\theta^{\dot\alpha}[X_{\beta\dot\alpha},\lambda_\alpha]\right)  \nonumber \\
&&+\frac{1}{g^2}{\rm tr}\left(\bar\chi\gamma^a(X^a-m^a)\chi+\sqrt{2}\,i\,\varepsilon_{\alpha\beta}\,\bar\chi\lambda_{\alpha}\Phi_\beta - \sqrt{2}\,i\,\varepsilon_{\alpha\beta}\,\bar\Phi^{\alpha}\bar\lambda_{\beta}\chi\right)\ .
\end{eqnarray}
Here the indices $a = 1,\dots,5$ and correspond to the directions transverse to the D4-brane, while $m^a$ are the components of the bare mass of the flavours corresponding to the positions of the D4-branes. Also, ${\rm Tr}$ denotes trace over the $U(N)$ gauge indices (over the colours), while ${\rm tr}$ denotes a trace over the flavours. 

Note that in this notation the adjoint fermions (the pure BFSS part) are represented by four eight-component Weyl fermions in six dimensions $\lambda^{\rho}$ and $\theta^{\dot\alpha}$ correspondingly of positive and negative chirality and satisfying the reality conditions (simplectic majorana):
\begin{eqnarray}\label{reality-f}
\lambda_\alpha =\varepsilon_{\alpha\beta}\,\lambda^{c\,\beta};~~~\theta_{\dot\alpha} =-\varepsilon_{\dot\alpha\dot\beta}\ ,\theta^{c\, \dot\beta}\ ,
\end{eqnarray}
where:
\begin{equation}\label{cc}
\psi^c \equiv C_6^{-1}\bar\psi^{T}\ .
\end{equation}
Our goal is to relate the BFSS model in nine-dimensional notation (\ref{BFSS-9D}) to the six dimensional notation presented in equations (\ref{Berkooz--Douglas}) and (\ref{Berkooz--Douglas_int}). The nine-dimensional Minkowski Lagrangian can be obtained by reducing equation (\ref{BFSS 10Mink}) in the basis (\ref{gammas}) using the Weyl ansatz (\ref{Weyl}):
\begin{equation}
{\cal L}_{\rm BFSS} = \frac{1}{g^2}{\rm Tr}\left(\frac{1}{2}(D_0 X^i)^2 +\frac{1}{4}[X^i,X^j]^2+\frac{i}{2}\psi^T C_{9}\,D_0\psi +\frac{1}{2}\psi^T C_{9}\,\gamma^i[X^i,\psi]\right)\ .
\end{equation}
Our strategy is to obtain the adjoint part of equations (\ref{Berkooz--Douglas}), (\ref{Berkooz--Douglas_int}) by reduction in an appropriate ten dimensional basis for the gamma matrices and then relate the two frames by a unitary transformation. Our starting point is the basis:
\begin{eqnarray}\label{tgammas}
&&\tilde\Gamma^\mu = -\tilde\gamma^\mu\otimes \hat\gamma^5\ , ~~~~{\rm for }~ \mu = 0,\dots ,5\ , \nonumber \\
&&\tilde\Gamma^{5+m} = 1_{8}\otimes \hat\gamma^m\ ,  ~~~~{\rm for }~ m = 1,\dots ,4\ , \nonumber \\
&&\tilde\Gamma^{11} = -\tilde\gamma^7\otimes \hat\gamma^5\ , \nonumber \\
&&C_{10} =C_{6}\otimes C_{4}\ ,
\end{eqnarray}
where:
\begin{eqnarray}\label{5gammas}
&&\hat\gamma^1 = 1_2\otimes \sigma_2 , \nonumber \\
&&\hat\gamma^2 = 1_2\otimes \sigma_3 , ~~~~~\hat\gamma^5 = \hat\gamma^1\hat\gamma^2\hat\gamma^3\hat\gamma^4= \sigma_3\otimes \sigma_1  , \nonumber \\
&&\hat\gamma^3 = \sigma_2\otimes \sigma_1 , ~~~~~C_4 =\sigma_1\otimes i\sigma_2\ , \nonumber \\
&&\hat\gamma^4 = \sigma_1\otimes \sigma_1 , \nonumber 
\end{eqnarray}
and the six dimensional gamma matrices $\tilde\gamma^{\mu}$ are related to the matrices $\tilde\gamma_E$ appearing in equation (\ref{9gammas}) via: 
\begin{equation}
\tilde\gamma^a=\tilde\gamma_E^a~~,~ a = 1,\dots ,5;~~~\tilde\gamma^0=-i\tilde\gamma_E^6;~~~\tilde\gamma^7=-\tilde\gamma_E^7\ .
\end{equation}
One can now easily check that the frames (\ref{gammas}) and (\ref{tgammas}) are related via the unitary transformation $S$:
\begin{eqnarray}\label{trasnformationS}
&&S\,\tilde\Gamma^M\,S^{-1}\,=\,\Gamma^M;~~~S=S_1 S_2; ~~[S_1,S_2]=0; \nonumber \\
&&S_1=\frac{1}{\sqrt{2}}(1+i\tilde\Gamma^0\tilde\Gamma^6)=\frac{1}{\sqrt{2}}(1+\tilde\gamma^0\otimes\sigma_3\otimes\sigma_3)\ , \nonumber\\
&&S_2=\frac{1}{\sqrt{2}}(1+\tilde\Gamma^7\tilde\Gamma^{11})=\frac{1}{\sqrt{2}}(1-\tilde\gamma^7\otimes\sigma_3\otimes i\sigma_2)\ .
\end{eqnarray}
Furthermore, one can check that the charge conjugation matrix $C_{10}$ is invariant under the transformation $S$, namely that: $S^TC_{10}\,S\,=\,C_{10}$.
Now let us focus on reducing the Lagrangian
\begin{equation}\label{tact}
{\cal L} =\frac{1}{g^2}{\rm Tr}\left(\frac{1}{2}({\cal D}_0X^i)^2 +\frac{1}{4}[X^i,X^j]^2-\frac{i}{2}\tilde\Psi^T C_{10}\,\tilde\Gamma^0D_0\tilde\Psi +\frac{1}{2}\tilde\Psi^T C_{10}\,\tilde\Gamma^i[X^i,\tilde\Psi]\right)
\end{equation}
in the basis (\ref{tgammas}), where $\tilde\Psi$ is a Majorana-Weyl spinor satisfying:
\begin{equation}\label{Mj-Weyl}
\tilde\Gamma^{11}\,\tilde\Psi \,=\,\tilde\Psi;~~~~{\tilde\Psi}^{\dagger}\,\tilde\Gamma^0 \,=\,\tilde\Psi^T\,C_{10}; \ .
\end{equation}
We consider the ansatz:
\begin{equation}\label{tPsi}
\tilde\Psi=\frac{1}{\sqrt{2}}\left(\begin{array}{c} \lambda^1 \\ \theta^{\dot1} \end{array}\right)\otimes \left(\begin{array}{c} 1 \\ -1 \end{array}\right)+\frac{1}{\sqrt{2}}\left(\begin{array}{c} \theta^{\dot2} \\ \lambda^{2} \end{array}\right)\otimes \left(\begin{array}{c} 1 \\ 1 \end{array}\right)\ .
\end{equation}
One can easily check that the Weyl condition on $\tilde\Psi$ (first relation in equation (\ref{Mj-Weyl})) implies:
\begin{eqnarray}
&&P_+\,\lambda^\rho \,=\,\lambda^\rho;~~~P_{-}\,\theta^{\dot\rho} \,=\,\theta^{\dot\rho}; \ , \nonumber \\
&&P_{\pm}\equiv\frac{1}{2}\left({1\pm\tilde\gamma^7}\right)\ .
\end{eqnarray}
The Majorana condition can be rewritten as: $\tilde\Psi = C_{10}^{-T}\tilde\Gamma^{0\, T}\tilde\Psi^* = -C_{10}^{-1}\tilde\Gamma^{0\, T}\tilde\Psi^*$, where we used that: $C_{10}^{T}=-C_{10}$. Using equations  (\ref{cc}) and (\ref{tgammas}) one obtains:
\begin{equation}\label{conjtPsi}
\tilde\Psi=-C_{10}^{-1}\tilde\Gamma^{0\, T}\tilde\Psi^*=\frac{1}{\sqrt{2}}\left(\begin{array}{c} \lambda^{c\,2} \\ -\theta^{c\,\dot2} \end{array}\right)\otimes \left(\begin{array}{c} 1 \\ -1 \end{array}\right)+\frac{1}{\sqrt{2}}\left(\begin{array}{c} \theta^{c\,\dot1} \\ -\lambda^{c\,1} \end{array}\right)\otimes \left(\begin{array}{c} 1 \\ 1 \end{array}\right)\ .
\end{equation}
Comparing equations (\ref{tPsi}) and (\ref{conjtPsi}) one arrives at the reality condition (\ref{reality-f}). 

Let us now reduce the Lagrangian (\ref{tact}). One easily obtains:
\begin{eqnarray}
&&-\frac{i}{2}\tilde\Psi^T C_{10}\,\tilde\Gamma^0D_0\tilde\Psi = \frac{i}{2}\lambda^{\dagger\,\rho}D_0\lambda_\rho+\frac{i}{2}\theta^{\dagger\dot\rho}D_0\theta_{\dot\rho}\ , \nonumber \\
&&\frac{1}{2}\tilde\Psi^T C_{10}\,\tilde\Gamma^a[X^a,\tilde\Psi] = \frac{1}{2}\bar\lambda^\rho\tilde\gamma^a[X^a,\lambda_\rho]+\frac{1}{2}\bar\theta^{\dot\alpha}\tilde\gamma^a[X^a,\theta_{\dot\alpha}]\, ~~~\ ,~a=1,\dots,5\ ,
\end{eqnarray}
which agree with the corresponding terms in equations (\ref{Berkooz--Douglas}) and (\ref{Berkooz--Douglas_int}). Reducing the last term in (\ref{tact}) along the directions parallel to the D4-branes is a bit more involved. We obtain:
\begin{eqnarray}
&&{\rm Tr}\left(\frac{1}{2}\tilde\Psi^T C_{10}\,\tilde\Gamma^6[X^6,\tilde\Psi] \right)= {\rm Tr}\left(i\bar\theta^{\,\dot1}[X^6,\lambda^2]-i\bar\theta^{\,\dot2}[X^6,\lambda^1]\right)\ , \nonumber \\
&&{\rm Tr}\left(\frac{1}{2}\tilde\Psi^T C_{10}\,\tilde\Gamma^7[X^7,\tilde\Psi] \right)= {\rm Tr}\left(-\bar\theta^{\dot1}[X^7,\lambda^2]-\bar\theta^{\,\dot2}[X^7,\lambda^1]\right)\ , \nonumber  \\
&&{\rm Tr}\left(\frac{1}{2}\tilde\Psi^T C_{10}\,\tilde\Gamma^8[X^8,\tilde\Psi] \right)= {\rm Tr}\left(i\bar\theta^{\dot1}[X^8,\lambda^1]+i\bar\theta^{\,\dot2}[X^8,\lambda^2]\right)\ , \nonumber \\
&&{\rm Tr}\left(\frac{1}{2}\tilde\Psi^T C_{10}\,\tilde\Gamma^9[X^9,\tilde\Psi] \right)= {\rm Tr}\left(\bar\theta^{\dot1}[X^9,\lambda^1]-\bar\theta^{\,\dot2}[X^9,\lambda^2]\right)\ .
\end{eqnarray}
Therefore we have:
\begin{eqnarray}\label{red-mess}
{\rm Tr}\left(\frac{1}{2}\tilde\Psi^T C_{10}\,\tilde\Gamma^m[X^m,\tilde\Psi] \right)&=& {\rm Tr}\left(i\bar\theta^{\,\dot1}[X^6+iX^7,\lambda^2]+i\bar\theta^{\,\dot2}[X^8+iX^9,\lambda^2]\right. \nonumber\\
&&~~\left. - i\bar\theta^{\dot1}[-X^8+iX^9,\lambda^1] -i\bar\theta^{\,\dot2}[X^6-iX^7,\lambda^1]\right)\\
\label{mess}
{\rm Tr}\left(-\sqrt{2}\,i\,\varepsilon_{\alpha\beta}\,\bar\theta^{\dot\alpha}[X_{\beta\dot\alpha},\lambda_\alpha]\right) &=&{\rm Tr}\left(i\bar\theta^{\,\dot1}[\sqrt{2}\,X_{1\dot1},\lambda^2]+i\bar\theta^{\,\dot2}[\sqrt{2}\,X_{1\dot2},\lambda^2]\right. \nonumber\\
&&~~\left. - i\bar\theta^{\dot1}[\sqrt{2}\,X_{2\dot1},\lambda^1] -i\bar\theta^{\,\dot2}[\sqrt{2}\,X_{2\dot2},\lambda^1]\right)\ .
\end{eqnarray}
Comparing the terms in equations (\ref{red-mess}) and (\ref{mess}) we conclude that\footnote{Note that our expression for $X_{\rho\dot\rho}$ differs from the one in ref.~\cite{VanRaamsdonk:2001cg} by the reflection $X^8\to-X^8$.}:
\begin{equation}\label{Xrhorho}
||X_{\rho\dot\rho}||=\frac{1}{\sqrt{2}}\left(\sigma_0\,X^6+i\sigma^A\,X^{10-A}\right) =\frac{1}{\sqrt{2}}\left(\begin{array}{cc} X^6+iX^7 &  X^8+iX^9\\ -X^8+iX^9 & X^6-iX^7 \end{array}\right)\ .
\end{equation}

Our next step is to express $\lambda^\rho$ and $\theta^{\dot\rho}$ in terms of the spinor field $\psi$ defined in equation (\ref{Weyl}). To this end one has to use the relation\footnote{One can check that by substituting $\tilde\Psi = S^{-1}\Psi$ and $C_{10}=S^TC_{10}S$ into equation (\ref{tact}) and using the transformation (\ref{trasnformationS}) one will arrive at equation (\ref{BFSS 10Mink}).} $\tilde\Psi = S^{-1}\Psi$. Decomposing:
\begin{equation}
\Psi= \left(\begin{array}{c} \psi_1 \\  \psi_2 \end{array}\right)\otimes \left(\begin{array}{c} 1 \\ 0 \end{array}\right)
\end{equation}
and using the definitions in equation (\ref{trasnformationS}) one arrives at:
\begin{eqnarray}
S_2^{\dagger}\Psi &=&\frac{1}{\sqrt{2}}\left(\begin{array}{c} \psi_1 \\ \psi_2 \end{array}\right)\otimes \left(\begin{array}{c} 1 \\ 0 \end{array}\right)-\frac{1}{\sqrt{2}}\left(\begin{array}{c} \tilde\gamma^7\psi_1 \\ -\tilde\gamma^7\psi_2 \end{array}\right)\otimes \left(\begin{array}{c} 0 \\ 1 \end{array}\right)\nonumber \\
&=&\frac{1}{\sqrt{2}}\left(\begin{array}{c} P_+\psi_1 \\ P_-\psi_2 \end{array}\right)\otimes \left(\begin{array}{c} 1 \\ -1 \end{array}\right)+\frac{1}{\sqrt{2}}\left(\begin{array}{c} P_-\psi_1 \\ P_+\psi_2 \end{array}\right)\otimes \left(\begin{array}{c} 1 \\ 1 \end{array}\right)\ ,\\
S_1^{\dagger}S_2^{\dagger}\Psi&=& \frac{1}{\sqrt{2}}\left(\begin{array}{c} P_+\,e^{-\frac{\pi}{4}\tilde\gamma^0}\,\psi_1 \\ P_-\,e^{\frac{\pi}{4}\tilde\gamma^0}\psi_2 \end{array}\right)\otimes \left(\begin{array}{c} 1 \\ -1 \end{array}\right)+\frac{1}{\sqrt{2}}\left(\begin{array}{c} P_-\,e^{-\frac{\pi}{4}\tilde\gamma^0}\psi_1 \\ P_+\,e^{\frac{\pi}{4}\tilde\gamma^0}\psi_2 \end{array}\right)\otimes \left(\begin{array}{c} 1 \\ 1 \end{array}\right),~\label{compare1}
\end{eqnarray}
where $P_{\pm} =\frac{1}{2}(1\pm\tilde\gamma^7)$. Comparing equations (\ref{compare1}) and (\ref{tPsi}) we arrive at:
\begin{eqnarray}\label{trans}
\lambda_1&=&P_+\,e^{-\frac{\pi}{4}\tilde\gamma^0}\,\psi_1\ , \nonumber \\
\lambda_2&=&P_+\,e^{\frac{\pi}{4}\tilde\gamma^0}\psi_2\ , \nonumber \\
\theta_{\dot1}&=&P_-\,e^{\frac{\pi}{4}\tilde\gamma^0}\psi_2\ , \nonumber \\
\theta_{\dot2}&=&P_-\,e^{-\frac{\pi}{4}\tilde\gamma^0}\psi_1\ . \nonumber 
\end{eqnarray}
Equations (\ref{Berkooz--Douglas}) and (\ref{Berkooz--Douglas_int}) can then be written as:
\begin{eqnarray}\label{BD-mixed}
{\cal L} &=& \frac{1}{g^2}{\rm Tr}\left(\frac{1}{2}D_0X^iD_0X^i+\frac{i}{2}\psi^T C_{9}\,D_0\psi\right)+\frac{1}{g^2}{\rm tr}\left(D_0\bar\Phi^{\rho}D_0\Phi_\rho+i\chi^{\dagger}D_0\chi\right)+{\cal L}_{\rm int}\ ,\nonumber \\
{\cal L}_{\rm int} &=&\frac{1}{g^2}{\rm Tr}\left(\frac{1}{4}[X^i,X^j]^2+\frac{1}{2}\psi^T C_{9}\,\gamma^i[X^i,\psi]\right)+\frac{1}{g^2}{\rm tr}\left(\bar\chi\gamma^a(X^a-m^a)\chi-\bar\Phi^{\rho}(X^a-m^a)^2\Phi_\rho\right) \nonumber \\
&&+\frac{1}{g^2}{\rm tr}\left(\bar\Phi^{\alpha}[\bar X^{\beta\dot\alpha},X_{\alpha\dot\alpha}]\Phi_\beta+\frac{1}{2}\bar\Phi^\alpha\Phi_\beta \bar\Phi^\beta\Phi_\alpha-\bar\Phi^\alpha\Phi_\alpha \bar\Phi^\beta\Phi_\beta\right) \nonumber \\
&&+\frac{1}{g^2}i\sqrt{2}\,{\rm tr}\left(\bar\Phi^2\,\bar\psi_1\,e^{\frac{\pi}{4}\tilde\gamma^0}\chi-\bar\Phi^1\,\bar\psi_2\,e^{-\frac{\pi}{4}\tilde\gamma^0}\chi+ \bar\chi\,e^{-\frac{\pi}{4}\tilde\gamma^0}\psi_1\,\Phi^2-\bar\chi\,e^{\frac{\pi}{4}\tilde\gamma^0}\psi_2\,\Phi^1\right)\ .
\end{eqnarray}
\subsubsection{The fundamental fermions. Wick rotation.}
Next we focus on wick rotating the action (\ref{BD-mixed}). Note that before wick rotating the fermions it is crucial to use the reality condition $\bar\psi =\psi^T\,C_{10}$, which implies $\psi_1^*=\psi_2\ , \psi_2^*=\psi_1$. To wick rotate the Weyl fermions $\chi$ it is convenient to first rewrite the action in five dimensional notation,  using five dimensional Dirac fermions. To this end we use an explicit basis for $\tilde\gamma^\mu$~\cite{Davies}: 
\begin{eqnarray}\label{majorana-gamma}
\tilde\gamma^0&=&-i\sigma_3\otimes C_5\ , ~~~~~~~~~~~\gamma'^{1}\,=\,-\sigma_2\otimes \sigma_1\ ,\nonumber \\
\tilde\gamma^1&=&-i\sigma_3\otimes C_5\,\gamma'^{1}\ , ~~~~~~~\gamma'^{2}\,=\,-\sigma_2\otimes \sigma_2\ ,\nonumber \\
\tilde\gamma^2&=&\sigma_1\otimes C_5\,\gamma'^{2}\ , ~~~~~~~~~~\,\gamma'^{3}\,=\,-\sigma_2\otimes \sigma_3\ ,\nonumber \\
\tilde\gamma^3&=&-i\sigma_3\otimes C_5\,\gamma'^{3}\ , ~~~~~~~\gamma'^{4}\,=\,\sigma_1\otimes 1_2\ ,\nonumber \\
\tilde\gamma^4&=&\sigma_1\otimes C_5\,\gamma'^{4}\ , ~~~~~~~~~~\,\gamma'^{5}\,=\,\sigma_3\otimes 1_2\ ,\nonumber \\
\tilde\gamma^5&=&\sigma_1\otimes C_5\,\gamma'^{5}\ , ~~~~~~~~~~~C_5=1_2\otimes \sigma_2\ ,\nonumber \\
\tilde\gamma^7&=&\sigma_2\otimes 1_4\ , \nonumber 
\end{eqnarray}
where $C_5$ is a charge conjugation matrix satisfying $C_5\,\gamma'^{m}\,C_5^{\,-1}\,=\,\gamma'^{m\,T}$. Using that $\lambda^\alpha$ and $\chi$ are of positive and negative chirality we define:
\begin{eqnarray}\label{gen-red}
\chi&=&\frac{1}{\sqrt{2}}\left(\begin{array}{c} 1 \\ -i \end{array}\right)\otimes \hat\chi\ , ~~~~~\lambda^\alpha=\frac{1}{\sqrt{2}}\left(\begin{array}{c} 1 \\ i \end{array}\right)\otimes \hat\lambda^\alpha\ , \nonumber \\ 
\psi_{\alpha}&=&\frac{1}{\sqrt{2}}\left(\begin{array}{c} 1 \\ i \end{array}\right)\otimes\hat\psi_{\alpha,+}+\frac{1}{\sqrt{2}}\left(\begin{array}{c} 1 \\ -i \end{array}\right)\otimes\hat\psi_{\alpha,-}\ ,
\end{eqnarray}
where the hat symbol $\,\hat{}\,$ denotes four dimensional Dirac fermions. The transition rules (\ref{trans}) imply the relation:
\begin{eqnarray}\label{hatlam-psi}
\hat\lambda^1 &=& \frac{1}{\sqrt{2}}\left(\hat\psi_{1,+}+i\,C_5\,\hat\psi_{1,-}\right)\ ,\nonumber \\
\hat\lambda^2 &=& \frac{1}{\sqrt{2}}\left(\hat\psi_{2,+}-i\,C_5\,\hat\psi_{2,-}\right)\ .
\end{eqnarray}
Now we can reduce the fermionic part of the lagrangian (\ref{BD-mixed}). We obtain:
\begin{eqnarray}\label{red-eucledean-chi}
{\cal L}_\chi =\frac{1}{g^2}{\rm tr}\left(i\hat\chi^{\dagger}D_0\chi-\hat\chi^{\dagger}\gamma'^a(X^a-m^a)\hat\chi+\sqrt{2}\,\varepsilon^{\alpha\beta}\hat\chi_i^{\dagger}\,C_5\,\hat\lambda^A_\alpha\,T_{ij}^A\Phi_{j\,\beta}\,+\,\sqrt{2}\,i\hat\chi_i^{T}\hat\lambda^A_\alpha\,\bar T_{ij}^A\bar\Phi_j^\alpha\right)\ ,
\end{eqnarray}
where $T^A_{ij}$ are generator of $SU(N)$, also we have used the reduced version of the reality condition (\ref{reality-f}) to solve for $\hat\lambda^*_\alpha$ in terms of $\hat\lambda_\alpha$. Note that after Wick rotation the fermionic fields $\hat\chi$ and $\hat\chi^{\dagger}$ become independent. We thus define $\zeta^T = (\hat\chi^T\, ,\, \hat\chi^{\dagger})$ and Wick rotate taking ${\cal L}_\chi^{\rm E}=-{\cal L}_\chi(t\to-i\tau)$:
\begin{equation}\label{red-euclidean}
{\cal L}_\chi^{\rm E}=\frac{1}{g^2}{\rm tr}\left(\zeta_2^TD_{\tau}\zeta_1+\zeta_2^T\gamma'^a(X^a-m^a)\zeta_1-\sqrt{2}\,\varepsilon^{\alpha\beta}\zeta_{2\,i}^T\,C_5\,\hat\lambda^A_\alpha\,T_{ij}^A\Phi_{j\,\beta}\,-\,\sqrt{2}\,i\zeta_{1\,i}^{T}\hat\lambda^A_\alpha\,\bar T_{ij}^A\bar\Phi_j^\alpha\right)\ ,
\end{equation}
which is our expression for the Wick rotated part of the action involving $\chi$. A comment about the symmetry of the action is in order. We expect that it should have unbroken global $SO(5)$ symmetry. To verify this we study the action of the six dimensional $SO(1,5)$ symmetry on the reduced fermions. One can easily verify that the $SO(5)$ generators $\Sigma'^{ab}$ associated to the basis $\gamma'^m$ are embedded in $SO(1,5)$ via:
\begin{eqnarray}
\tilde \Sigma^{12}&=&-\sigma_2\otimes\Sigma'^{12}\ ,~~~\tilde \Sigma^{23}=-\sigma_2\otimes\Sigma'^{23}\ ,~~~\tilde \Sigma^{34}=-\sigma_2\otimes\Sigma'^{34}\ , \tilde \Sigma^{45}=1_2\otimes\Sigma'^{45}\ ,\nonumber \\
\tilde \Sigma^{13}&=&1_2\otimes\Sigma'^{13}\ , ~~~~~\,\tilde \Sigma^{24}=1_2\otimes\Sigma'^{24}\ ,~~~~\,~\tilde \Sigma^{35}=-\sigma_2\otimes\Sigma'^{35}\ ,\nonumber \\
\tilde \Sigma^{14}&=&-\sigma_2\otimes\Sigma'^{14}\ ,~~~\tilde \Sigma^{25}= 1_2\otimes\Sigma'^{25}\ , \nonumber \\
\tilde \Sigma^{15}&=&-\sigma_2\otimes\Sigma'^{15}\ ,
\end{eqnarray}
where $\tilde\Sigma^{ab}$ are the generators in the basis $\tilde\gamma^\mu$. One can also check that:
\begin{eqnarray}
\tilde\Sigma^{ab}\,\chi&=&\frac{1}{\sqrt{2}}\left(\begin{array}{c} 1 \\ -i \end{array}\right)\otimes \Sigma'^{ab}\hat\chi\ \ ,\nonumber \\
\tilde\Sigma^{ab}\,\lambda_\alpha&=&\frac{1}{\sqrt{2}}\left(\begin{array}{c} 1 \\ i \end{array}\right)\otimes \left(-\Sigma'^{ab\,T}\right)\hat\lambda_\alpha \ .
\end{eqnarray}
We see that: $\zeta_1$ transforms as $\hat\chi$, $\zeta_2^T$ transform as $\hat\chi^{\dagger}$ and $\hat\lambda_\alpha$ transforms as $C_5\,\chi$, which ensures the $SO(5)$ invariance of the reduced action. We can make this explicit by defining $\zeta'^T=(\hat\chi^T\,,\,\hat\chi^{\dagger}\,C_5)$ and $\hat\lambda'_\alpha = C_5\lambda_\alpha$. Now all the fields $\zeta'_1$, $\zeta'_2$ and $\hat\lambda'_\alpha$ transform as $\hat\chi$ and the lagrangian takes the form:
\begin{eqnarray}
{\cal L}_\chi^{\rm E}&=&\frac{1}{g^2}{\rm tr}\left(\zeta'^T_2\,C_5D_{\tau}\zeta'_1+\zeta'^T_2C_5\gamma'^a(X^a-m^a)\zeta_1\right. \nonumber \\
&&\left.-\sqrt{2}\,\varepsilon^{\alpha\beta}\zeta'^T_{2\,i}\,C_5\,\hat\lambda'^A_\alpha\,T_{ij}^A\Phi_{j\,\beta}\,-\,\sqrt{2}\,i\zeta'^T_{1\,i}C_5\hat\lambda'^A_\alpha\,\bar T_{ij}^A\bar\Phi_j^\alpha\right)\ ,
\end{eqnarray}
 which is explicitly $SO(5)$ invariant. For technical reasons we will keep the non-standard form of the lagrnagian (\ref{red-euclidean}).

 \subsubsection{The fundamental fermions. Discretisation.}
 Next we focus on discretising the action corresponding to (\ref{red-euclidean}):
 \begin{equation}
S_{\chi}^{\rm E}=\frac{1}{g^2}\int\limits_0^\beta d\tau {\rm tr}\left(\zeta_2^TD_{\tau}\zeta_1+\zeta_2^T\gamma'^a(X^a-m^a)\zeta_1-\sqrt{2}\,\varepsilon^{\alpha\beta}\zeta_{2\,i}^T\,C_5\,\hat\lambda^A_\alpha\,T_{ij}^A\Phi_{j\,\beta}\,-\,\sqrt{2}\,i\zeta_{1\,i}^{T}\hat\lambda^A_\alpha\,\bar T_{ij}^A\bar\Phi_j^\alpha\right)
 \end{equation}
 Using the link variables (\ref{linkU}) for the covariant derivative $D_\tau \zeta_1$ we can write
 \begin{equation}
D_\tau\,\zeta^1\to\frac{U_{n,n+1}\zeta^1_{n+1}-\zeta^1_n}{a}\ .
 \end{equation}
 Using again the gauge in which the holonomy is concentrated at one link (see Appendix B) and imposing anti-periodic boundary conditions on the fermions, for the kinetic term we obtain:
 \begin{equation}
\int\limits_0^\beta d\tau\,{\rm tr}\,\left(\zeta_2^T\,D_\tau\,\zeta_1\right)={\rm tr}\left(\sum\limits_{n=0}^{\Lambda-2}\zeta_n^{2\,T}\,\zeta_{n+1}^{1}-\zeta_{\Lambda-1}^{2\,T}\,D\,\zeta_0^1-\sum\limits_{n=0}^{\Lambda-1}\zeta_n^{2\,T}\,\zeta_n^1\right)\ ,
 \end{equation}
where $D ={\rm diag}\{e^{i\theta_1},\dots,e^{i\theta_N}\}$ is the holonomy matrix. Defining the matrix:
\begin{equation}
K_{\chi\,n,m}^{ij}=\delta_{n+1,m}\,\delta_{ij}-\delta_{n,\Lambda-1}\,\delta_{m,0}\,D_{ij}-\delta_{n,m}\ ,
\end{equation}
 we can write:
 \begin{equation}
S_{\chi\,{\rm kin}}^{\rm E}=\frac{1}{g^2}\int\limits_0^\beta d\tau\,{\rm tr}\,\left(\zeta_2^T\,D_\tau\,\zeta_1\right)=\frac{1}{2g^2}\left(\zeta^{1\,T}\,,\,\zeta^{2\,T}\right)\left(\begin{array}{cc} 0_4 &  - K_\chi^T\\ K_\chi & 0_4 \end{array}\right)\left(\begin{array}{c} \zeta^1\\ \zeta^2 \end{array}\right)\ ,
 \end{equation}
 where we have suppressed all indices. One can show that the off-diagonal form of the kinetic term suppresses the fermion doubling, in the same way as the off diagonal choice of the charge conjugation matrix $C_9$ suppressed them for the adjoint fermions.Similarly for the potential term we get:
 \begin{equation}
\frac{1}{g^2}\int\limits_0^\beta d\tau\,{\rm tr}\,\left(\zeta_2^T\gamma'^a(X^a-m^a)\zeta_1\right)=\frac{a}{2g^2}\sum_{n=0}^{\Lambda-1}\left(\zeta^{1\,T}_n\,,\,\zeta^{2\,T}_n\right)\left(\begin{array}{cc} 0_4 &  - \bar\gamma'^a(\bar X_n^a-m^a)\\ \gamma'^a(X_n^a-m^a) & 0_4 \end{array}\right)\left(\begin{array}{c} \zeta^1_n\\ \zeta^2_n \end{array}\right)
 \end{equation}
 and
 \begin{eqnarray}
\frac{-\sqrt{2}}{g^2}\int\limits_0^\beta d\tau\,{\rm tr}\,\left(\varepsilon^{\alpha\beta}\zeta_{2\,i}^T\,C_5\,\hat\lambda^A_\alpha\,T_{ij}^A\Phi_{j\,\beta}\,+i\zeta_{1\,i}^{T}\hat\lambda^A_\alpha\,\bar T_{ij}^A\bar\Phi_j^\alpha\right)&=&\sum_{n=0}^{\Lambda-1}{\rm tr}\left(\zeta_{n\,i}^T\,\left({\cal M}_{\zeta\,\lambda}\right)^{A}_{n\,i}\hat\lambda_n^A\right. \nonumber \\
&& \left.-\hat\lambda^{A\,T}_n\,\left({\cal M}_{\zeta\,\lambda}^{T}\right)^A_{n\,i}\zeta_{n\,i} \right)\ ,
 \end{eqnarray}
 where $\hat\lambda^A_n\equiv(\hat\lambda^A_{1\,n}\,,\,\hat\lambda^A_{2\,,\,n})$ and:
 \begin{equation}
\left({\cal M}_{\zeta\lambda}\right)_{n\,i}^A =-\frac{\sqrt{2}\,a}{2g^2}\left(\begin{array}{cc} i\,\bar T_{ij}^A\,\bar\Phi^1_{j\,n} & i\,\bar T_{ij}^A\,\bar\Phi^2_{j\,n}\\  C_5\,T_{ij}^A\,\Phi^2_{j\,n} & -C_5\,T_{ij}^A\,\Phi^1_{j\,n} \end{array}\right)\ .
 \end{equation}
 Altogether we can write:
 \begin{equation}
 S_{\chi}^{\rm E}={\rm tr}\left(\zeta^T\,{\cal M}_{\zeta\zeta}\,\zeta+\zeta^T\,{\cal M}_{\zeta\lambda}\,\hat\lambda-\hat\lambda^T\,{\cal M}_{\zeta\lambda}^T\,\zeta\right)\ ,
 \end{equation}
 where we have suppressed all of the indices and $M_{\zeta\zeta}$ is given by:
 \begin{equation}
M_{\zeta\zeta\,n,m}^{ij}=\frac{1}{2g^2}\left(\begin{array}{cc} 0_4 & -K_{\chi\,m,n}^{ji}-a\,{\bar\gamma}'^a(\bar X_{n\,ij}^a-m^a\delta_{ij})\,\delta_{n,m}\\  K_{\chi\,n,m}^{ij}+a\,\gamma'^a(X_{n\,ij}^a-m^a\delta_{ij})\,\delta_{n,m} & 0_4 \end{array}\right)
 \end{equation}
 One should keep in mind that $\hat\lambda$ can be expressed in terms of $\psi$, namely $\hat\lambda = {\cal M}_{\lambda\psi}\psi$. More explicitly using equations (\ref{gen-red}) and (\ref{hatlam-psi}) we obtain:
 \begin{equation}
\left(\begin{array}{c} \hat\lambda^A_{1\,n} \\ \hat\lambda^A_{2\,n} \end{array}\right)=\frac{1}{\sqrt{2}}\left(\begin{array}{cccc} e^{i\frac{\pi}{4}C_5} & -i\, e^{-i\frac{\pi}{4}C_5} & 0_4 & 0_4\\ 0_4 & 0_4 & e^{-i\frac{\pi}{4}C_5} & -i\,e^{i\frac{\pi}{4}C_5}\end{array}\right)\left(\begin{array}{c} \psi^A_{1\,n} \\ \psi^A_{2\,n} \end{array}\right)\ .
 \end{equation}
 Clearly, defining ${\cal M}_{\zeta\psi}={\cal M}_{\zeta\lambda}\,{\cal M}_{\lambda\psi}$ we can write:
 \begin{equation}
 S_{\chi}^{\rm E}={\rm tr}\left(\zeta^T\,{\cal M}_{\zeta\zeta}\,\zeta+\zeta^T\,{\cal M}_{\zeta\psi}\,\psi-\psi^T\,{\cal M}_{\zeta\psi}^T\,\zeta\right)\ ,
 \end{equation}
  Finally, using equations (\ref{M-psi}) and (\ref{S-psi}) for the total fermionic action we obtain:
  \begin{equation}
S_f^{\rm tot} =(\psi^T\,,\,\zeta^T)\left(\begin{array}{cc} {\cal M} & -{\cal M}^T_{\zeta\psi}\\  {\cal M}_{\zeta\psi} & {\cal M}_{\zeta\zeta} \end{array}\right)\left(\begin{array}{c} \psi\\  \zeta  \end{array}\right) = (\psi^T\,,\,\zeta^T)\,{\cal M}_{\rm  tot}\,\left(\begin{array}{c} \psi\\  \zeta  \end{array}\right)\ .
  \end{equation}
  \subsubsection{RHMC and pseudo-fermionic forces}
  The next step is to apply the RHMC method \cite{Clark:2004cp} to the model. To this end we need the so called pseudo-fermionic forces. Let us summarise briefly the philosophy. The partition function of the model can be written as:
  \begin{equation}
{\cal Z} \propto\int{\cal D}X\,{\cal D}\bar\Phi\,{\cal D}\Phi\,e^{-S_{\rm bos}[X,\Phi]-S_f^{\rm tot}}\propto \int{\cal D}X\,{\cal D}\bar\Phi\,{\cal D}\Phi\,{\rm Pf}({\cal M}_{\rm tot})\,e^{-S_{\rm bos}[X,\Phi]}
  \end{equation}
  Assuming that the model does not suffer from a severe sign problem we can ignore the phase of the Pfaffian and use that:
  \begin{equation}
|{\rm Pf}({\cal M}_{\rm tot})|={\rm det}({\cal M}_{\rm tot}^{\dagger}\,{\cal M}_{\rm tot})^{1/4}\ ,
  \end{equation}
  to write 
  \begin{equation}
{\cal Z} \propto \int{\cal D}X\,{\cal D}\bar\Phi\,{\cal D}\Phi\,{\cal D}\xi^{\dagger}\,{\cal D}\xi\,e^{-S_{\rm bos}[X,\Phi]-S_{\rm ps.f}}\ ,
  \end{equation}
  where
  \begin{equation}
S_{\rm ps.f} \equiv\xi^{\dagger}\,({\cal M}_{\rm tot}^{\dagger}\,{\cal M}_{\rm tot})^{-1/4}\xi\ .
  \end{equation}
  Here $\xi$ is a $16(N_c^2-1)\Lambda+8\,N_f\,Nc\,\Lambda$ dimensional vector consisting of the pseudo-fermionic fields. The idea of the RHMC is to approximate the rational exponent of the matrix ${\cal M}_{\rm tot}^{\dagger}\,{\cal M}_{\rm tot}$ with a partial sum:
  \begin{equation}\label{partial}
({\cal M}_{\rm tot}^{\dagger}\,{\cal M}_{\rm tot})^{\delta} =\alpha_0+\sum_{i=1}^{\#}\alpha_i\,({\cal M}_{\rm tot}^{\dagger}\,{\cal M}_{\rm tot}+\beta_i)^{-1}\ ,
  \end{equation}
  where the parameters $\alpha_0,\alpha_i,\beta_i$ and $\#$  depend on the rational exponent $\delta$, the spectral range of the matrix ${\cal M}_{\rm tot}^{\dagger}\,{\cal M}_{\rm tot}$ and the desired accuracy. We will need two rational exponents. To update the pseudo fermions we use that the field $\eta \equiv ({\cal M}_{\rm tot}^{\dagger}\,{\cal M}_{\rm tot})^{-1/8}\xi$ has a gaussian distribution and solve for $\xi = ({\cal M}_{\rm tot}^{\dagger}\,{\cal M}_{\rm tot})^{1/8}\,\eta$ using a multi-shift solver. Therefore, $\delta =1/8$ is one of the rational exponents that we need. To calculate the fermionic forces and the contribution to the hamiltonian we need to invert $({\cal M}_{\rm tot}^{\dagger}\,{\cal M}_{\rm tot})^{-1/4}$ and the second exponent is $\delta =-1/4$. 
  
  Let us elaborate on the computation of the fermionic forces. We have three type of forces: derivative with respect to $X_{n\,ij}$, derivatives with respect to $\bar\Phi^{\alpha}_{n\,i}$ and derivative with respect to the phases of the links $\theta_i$. Using the partial expansion (\ref{partial}), one can easily derive expression for the derivatives of the fermionic action:
  \begin{eqnarray}\label{ps.ferm-force}
\frac{\partial S_{\rm ps.f}}{\partial u} &=&-\sum_{i=1}^{\#}\alpha_i\,\xi^{\dagger}({\cal M}_{\rm tot}^{\dagger}\,{\cal M}_{\rm tot}+\beta_i)^{-1}\,\frac{\partial({\cal M}_{\rm tot}^{\dagger}\,{\cal M}_{\rm tot})}{\partial u}\,({\cal M}_{\rm tot}^{\dagger}\,{\cal M}_{\rm tot}+\beta_i)^{-1}\xi \nonumber \\
&&=-\sum_{i=1}^{\#}\alpha_i\,h_i^{\dagger}\,\frac{\partial({\cal M}_{\rm tot}^{\dagger}\,{\cal M}_{\rm tot})}{\partial u}\,h_i\ ,
  \end{eqnarray}
  where $h_i$ satisfy $({\cal M}_{\rm tot}^{\dagger}\,{\cal M}_{\rm tot}+\beta_i)\,h_i=\xi_i$ and are obtained from the multi-solver. 
  
  \subsubsection{Bosonic action. Discretisation}
  
  Finally, we focus on the bosonic action of the fundamental fields. Wick rotating the action for $\Phi$, we obtain:
  \begin{equation}\label{action-phi}
{\cal S}^{\rm E}_{\Phi}=\frac{1}{g^2}\int\limits_0^\beta\,d\tau\,{\rm tr}\left(D_\tau\bar\Phi^{\rho}D_\tau\Phi_\rho  -\bar\Phi^{\alpha}[\bar X^{\beta\dot\alpha},X_{\alpha\dot\alpha}]\Phi_\beta+\bar\Phi^{\rho}(X^a-m^a)^2\Phi_\rho-\frac{1}{2}\bar\Phi^\alpha\Phi_\beta \bar\Phi^\beta\Phi_\alpha+\bar\Phi^\alpha\Phi_\alpha \bar\Phi^\beta\Phi_\beta \right)\ .
  \end{equation}
 Before we discretise the action (\ref{action-phi}) it is instructive to massage the term $\bar\Phi^{\alpha}[\bar X^{\beta\dot\alpha},X_{\alpha\dot\alpha}]\Phi_\beta$. First we point out that \cite{VanRaamsdonk:2001cg}:
 \begin{equation}
\bar X^{\sigma\dot\sigma}=\varepsilon^{\sigma\rho}\,\varepsilon^{\dot\sigma\dot\rho}\,X_{\rho\dot\rho}\ ,
 \end{equation}
  which in matrix notation becomes $\bar X =\sigma_2\,X\sigma_2$. Next we define:
  \begin{eqnarray}
   \sigma^4\equiv -i\sigma_0\ ,~~~~Y^m\equiv X^{10-m}
  \end{eqnarray} 
   and rewrite equation (\ref{Xrhorho}) as:
  \begin{equation}
 || X_{\rho\dot\rho}||=\frac{1}{\sqrt{2}}\sum_{m=1}^4i\,\sigma^m Y^{m}\ ,
  \end{equation}
  while:
  \begin{equation}
||\bar X^{\sigma\dot\sigma}||=\frac{1}{\sqrt{2}}\sum_{m=1}^4i\,\sigma_2\sigma^m\sigma_2 Y^{m}\ ,
  \end{equation}
  We then obtain:
  \begin{equation}\label{comX}
[\bar X^{\beta\dot\alpha},X_{\alpha\dot\alpha}]=\frac{1}{2}(\sigma^m\sigma_2(\sigma^{l})^T\sigma_2)_{\alpha}^{\,\,\beta}\,[Y^m,Y^l]=-i(\sigma^A)_{\alpha}^{\,\,\beta}\left([Y^A,Y^4]+\frac{1}{2}\varepsilon^{ABC}[Y^B,Y^C]\right)\ .
  \end{equation}
  The last term in the equation above has a very clean group theory interpretation. To uncover it let us consider the following basis of the $SO(4)$ algebra:
  \begin{equation}\label{basis-L}
(L_{ab})_{cd}=i(\delta_{ad}\delta_{bc}-\delta_{ac}\delta_{bd})
  \end{equation}
  satisfying:
  \begin{equation}
[L_{ab},L_{cd}]=i(\delta_{ac}L_{bd}+\delta_{bd}L_{ac}-\delta_{ad}L_{bc}-\delta_{bc}L_{ad})\ .
  \end{equation}
The $SO(4)$ algebra can be split into $SU(2)\times SU(2)$ by considering the generators:
\begin{eqnarray}\label{JK}
J^A=\frac{1}{2}L_{A\,4}+\frac{1}{4}\varepsilon^{ABC}L_{BC}\ , ~~~~K^A=-\frac{1}{2}L_{A\,4}+\frac{1}{4}\varepsilon^{ABC}L_{BC} \ ~~~ A,B,C=1,\dots,3\ ,
\end{eqnarray}
  which satisfy:
  \begin{equation}
[J^A,J^B]=i\,\varepsilon^{ABC}J^C\ ,~~~[K^A,K^B]=i\,\varepsilon^{ABC}K^C\ ,~~~[J^A,K^A]=0\ .
  \end{equation}
  Now we notice that in the basis (\ref{basis-L}) we can write:
  \begin{equation}
  [Y^m,Y^n]=\frac{1}{2}i(L_{mn})_{ab}[Y^a,Y^b]\ .
  \end{equation}
  Substituting in equation (\ref{comX}) and using the definition of $J^A$ in equation (\ref{JK}) we obtain\footnote{Note that $J^A$ satisfy: $J^A_{ab}J^A_{cd}=\frac{1}{4}(\delta_{ac}\delta_{bd}-\delta_{ad}\delta_{bc})-\frac{1}{4}\varepsilon_{abcd}$ and aslo $J^A_{ab}J^B_{ba}=\delta^{AB}$.}:
  \begin{equation}
[\bar X^{\beta\dot\alpha},X_{\alpha\dot\alpha}]=(\sigma^A)_{\alpha}^{\,\,\beta}\,J^A_{ab}\,[Y^a,Y^b]
  \end{equation}
  We see that a general $SO(4)$ rotation acting on the $Y$'s would result on a $SO(3)$ rotation of $(\sigma^A)$, which would result in a $SU(2)$ rotation of $\bar\Phi^\alpha$ and $\Phi_\beta$ corresponding to the $(\frac12,0)$ representation of $SO(4)$. Finally defining 
\begin{equation}
J^{ab}_{\Phi\,ji} \equiv{\rm tr}(\bar\Phi_i^\alpha (\sigma^A)_{\alpha}^{\,\,\beta}\Phi_{\beta\,j})\,J^A_{ab}\ ,
\end{equation}
we can write:
\begin{equation}
-{\rm tr}\left(\bar\Phi^{\alpha}[\bar X^{\beta\dot\alpha},X_{\alpha\dot\alpha}]\Phi_\beta\right)={\rm Tr}\left(J_\Phi^{ab}\,[Y^b,Y^a]\right)\ ,
\end{equation}
while in (\ref{action-phi}) the term quartic in $\Phi$ becomes:
\begin{equation}
{\rm tr}\left(-\frac{1}{2}\bar\Phi^\alpha\Phi_\beta \bar\Phi^\beta\Phi_\alpha+\bar\Phi^\alpha\Phi_\alpha \bar\Phi^\beta\Phi_\beta \right)=\frac{1}{2}{\rm Tr}\left(J^{ab}_\Phi\,J^{ba}_\Phi\right)\ .
\end{equation}
The action (\ref{action-phi}) can then be written as:
\begin{equation}\label{actionPhi-J}
{\cal S}^{\rm E}_{\Phi}=\frac{1}{g^2}\int\limits_0^\beta\,d\tau\,\left\{{\rm tr}\left(D_\tau\bar\Phi^{\rho}D_\tau\Phi_\rho +\bar\Phi^{\rho}(X^a-m^a)^2\Phi_\rho\right)+{\rm Tr}\left(\frac{1}{2}J^{ab}_\Phi\,J^{ba}_\Phi+J_\Phi^{ab}\,[Y^b,Y^a]\right)\right\}\ .
\end{equation}
The discretisation of the action (\ref{actionPhi-J}) is straightforward. Using the link variables (\ref{linkU}) we define:
\begin{eqnarray}
D_{\tau}\Phi_{n,\alpha}&=&\frac{U_{n,n+1}\Phi_{n+1,\alpha}-\Phi_{n,\alpha}}{a}\nonumber \\
D_{\tau}\bar\Phi_{n}^{\alpha}&=&\frac{\bar\Phi_{n+1}^{\alpha}U_{n+1,n}-\bar\Phi_{n}^{\alpha}}{a}\ ,
\end{eqnarray}
where $a$ is the lattice spacing. Noting that $J^{ab}_{\Phi,n}$ transform the same way as $X_n$ under the gauge transformation (\ref{gauge-tr}), while $\Phi_{n,\alpha}$ transforms as $\Phi_{n,\alpha}\to (U_{0,1}\dots U_{n-1,n})\,\Phi_{n,\alpha}$ and using the gauge where the holonomy is concentrated at the $(0,\Lambda)$ link (see Appendix B), for the discrete action we obtain:
\begin{eqnarray}\label{flavour-bos-lattice}
{\cal S}_{\Phi} &=&-\frac{2N}{a}{\cal R}e\,{\rm tr}\left\{\sum\limits_{n=0}^{\Lambda-2}\bar\Phi^{\rho}_n\,\Phi_{n+1,\,\rho}+\bar\Phi^{\rho}_n\,D\,\Phi_{n+1,\,\rho}\right\}+N\,a\,\sum\limits_{n=0}^{\Lambda-1}\left\{{\rm tr}\left(\bar\Phi^{\rho}_n(X^a_n-m^a)^2\Phi_{n\,\rho}\right)+ \right. \nonumber\\
&&+\left. {\rm Tr}\left(\frac{1}{2}J^{ab}_{\Phi\,n}\,J^{ba}_{\Phi\,n}+J_{\Phi\,n}^{ab}\,[Y^b_n,Y^a_n]\right)\right\}\ , 
\end{eqnarray}
where $D ={\rm diag}\{e^{i\theta_1},\dots,e^{i\theta_N}\}$ is the holonomy matrix and without loss of generality we have set $g=1/\sqrt{N}$.

\section{Testing the correspondence}
\label{section test}
\begin{figure}[t] 
   \centering
      \includegraphics[width=3.in]{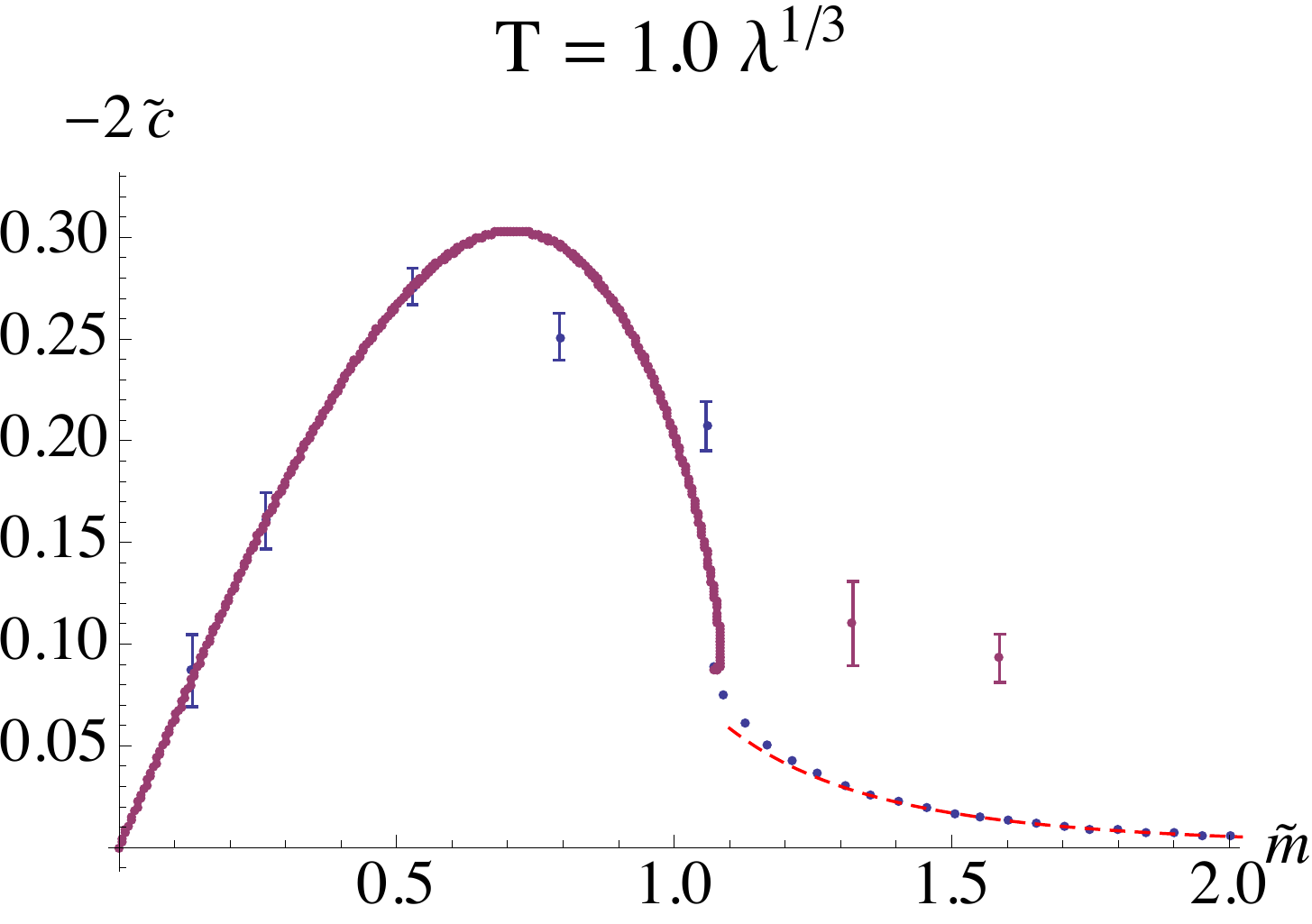} 
   \includegraphics[width=3.in]{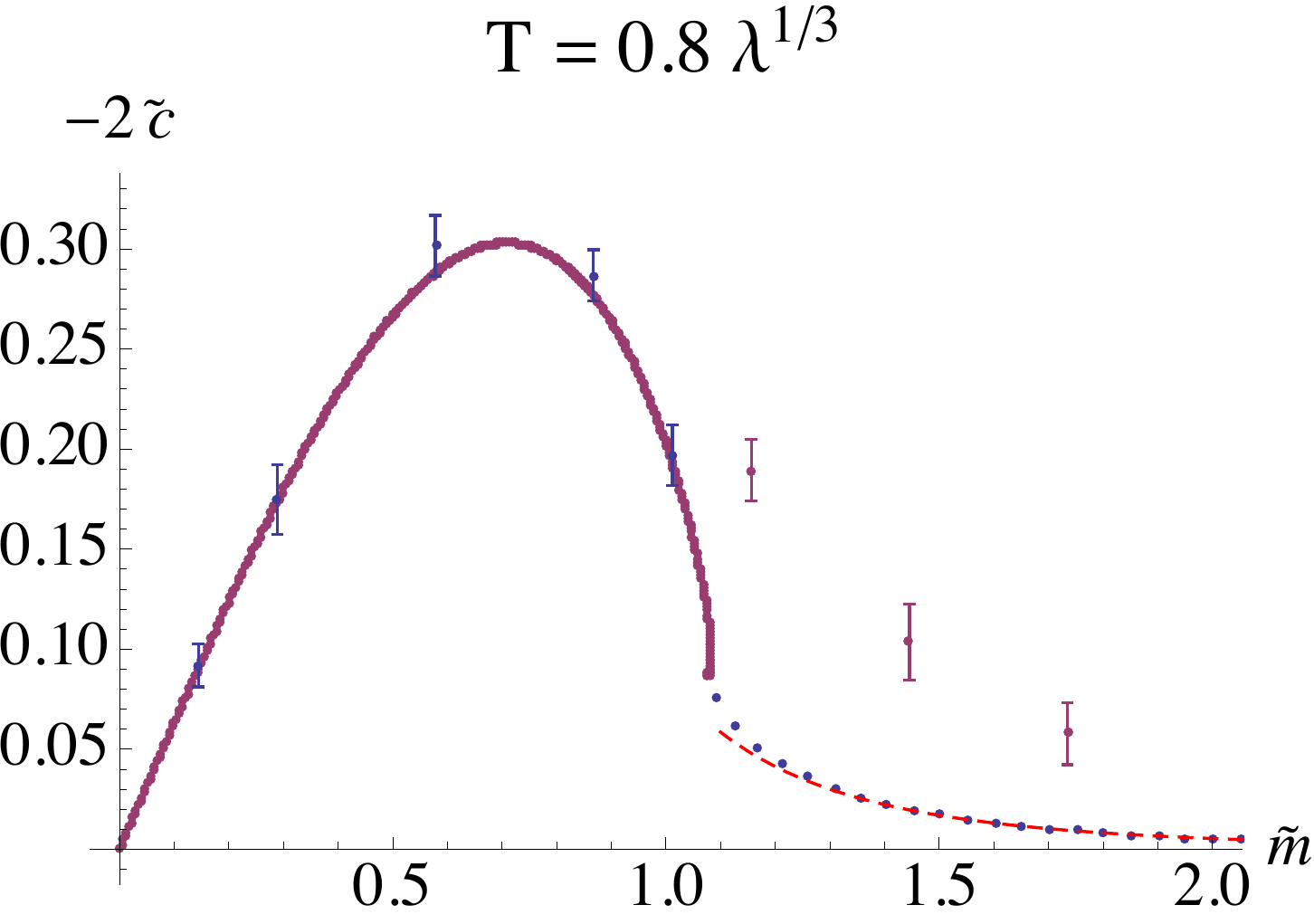} 
 \caption{\small Plots of the condensate versus bare mass parameter curve for $N=10$, $\Lambda =16$ and two different temperatures . {\it left:} For temperature $T = 1.0\,\lambda^{1/3}$ the curve shows excellent agreement at small masses, but deviates quickly from the theoretical curve at greater masses. {\it right:} The curve at temperature $T = 0.8\,\lambda^{1/3}$ exhibits excellent agreement throughout the whole range of masses corresponding to the black hole phase (blue error bars). Similarly to the higher temperature curve there is a significant deviation from the theoretical curve for the Minkowski phase (red error bars).} 
   \label{fig:3}
\end{figure}

In this section we compare the result of the lattice simulations of the model to the predictions of gauge gravity duality. Our main focus is the fundamental condensate of the theory. As definition of the condensate we use the derivative of the free energy of the theory with respect to the bare mass parameter $m^a$:
\begin{equation}
\langle {\cal O}_m^a\rangle \equiv \frac{\delta F}{\delta m^a} = \frac{1}{\beta}\left\langle\frac{\delta S_E}{\delta m^a} \right\rangle\ ,
\end{equation}
using equations (\ref{action-phi}) and (\ref{red-eucledean-chi}) for the condensate operator ${\cal O}_m$ we obtain:
\begin{equation}\label{Om}
 {\cal O}_m^a = \frac{1}{\beta\,g^2}\int\limits_0^\beta {\rm tr}\left(2\,\bar\Phi^\rho\,(m^a-X^a)\,\Phi_\rho +\hat\chi^{\dagger}\,\gamma^a\,\hat\chi\right)
\end{equation}
Using equation (\ref{flavour-bos-lattice}) it is straightforward to write down the discrete version of the bosonic term in (\ref{Om}), however this is not the case for the last term, since the fermions are not explicitly simulated on the lattice. The natural approach is to substitute the fermionic term with the derivative of the pseudo-fermionic action with respect to the mass parameter $m^a$. We obtain:
\begin{equation}
 {\cal O}_m^a = \frac{N}{\Lambda}\,\sum\limits_{n=0}^{\Lambda-1}{\rm tr}\left(2\,\bar\Phi^{\rho}_n\,(m^a-X^a_n)\,\Phi_{n\,\rho}\right)+\frac{\partial {S_{\rm ps. f}}}{\partial{m^a}}\ ,
\end{equation}
where the last term can be calculated by substituting the parameter $u$ in equation (\ref{ps.ferm-force}) with the mass parameter $m^a$. Note that we have also set $g^2 =1/N$, this implies that all dimensionful fields have been rescaled by an appropriate power of the 'tHooft coupling $\lambda = N\,g^2$. In particular the mass parameter $m^a$ has been rescaled by $\lambda^{1/3}$. The relation to the physical bare mass parameter appearing in equation (\ref{dictionary}) is then:
\begin{equation}
m^a = \frac{m_q}{\lambda^{1/3}}\,n^a\ ,
\end{equation}
where $n^a$ is a unit five-vector. Now using equation (\ref{dictionary}) the holographic prediction for the BD--parameters can be written as\footnote{Note that the condensate operator $ {\cal O}_m^a$ is dimensionless in $1+0$ dimensions and no scaling is needed.}:
\begin{eqnarray}
m^a &=&\left(\frac{120\,\pi^2}{49}\right)^{1/5}{\tilde T}^{\,2/5}\tilde m\,n^a\ , \nonumber \\
\langle {\cal O}_m^a\rangle &=&\left(\frac{2^4 \,15^3\,\pi^6}{7^6}\right)^{1/5}\,N_f\,N_c\,{\tilde T}^{\,6/5}\,(-2\,\tilde c)\,n^a\ ,
\label{dictionary-lattice}
\end{eqnarray}
where we have defined the dimensionless temperature $\tilde T = \frac{T}{\lambda^{1/3}}$, which is the parameter entering in the computer simulation (via $\tilde T^{-1} = \Lambda \,a$).  For simplicity we will use $n^a =(1,0,0,0,0)$. Equations (\ref{dictionary-lattice}) can then be used to scale the plot of $\langle {\cal O}_m^a\rangle$ versus $m^a$ obtained from computer simulations and compare to the $-2\,\tilde c$ versus $\tilde m$ plot presented in figure \ref{fig:2}. The resulting plots for two temperatures are presented in figure \ref{fig:3}. The plots are for matrix size $N=10$ and lattice spacing $\Lambda = 16$.

The left plot corresponds to temperature $T = 1.0\,\lambda^{1/3}$. One can observe excellent agreement between the gauge gravity duality and lattice simulations at small masses ($\tilde m < 1$). However, for greater masses there is a significant deviation from the theoretical curve. The right plot corresponds to temperature  $T = 0.8\,\lambda^{1/3}$. The excellent agreement between gauge/gravity predictions and lattice simulations extends for the whole range of masses within the deconfined (black hole) phase (blue error bars). In the deconfined (Minkowski) phase there are still significant deviations from the theoretical curve. These results are consistent with our discussion is section \ref{holo-condensate}, where we argued that the $\alpha'$ corrections to the supergravity background affect the black hole and Minkowski D4-brane embeddings differently. All black hole embeddings reach the horizon and as a result experience similar curvature effects for different values of the mass parameter therefore, the $\alpha'$ corrections largely cancel when one takes a derivative with respect to the mass to calculate the condensate. In contrast, Minkowski embeddings close at different radial distances above the horizon depending on the mass parameter. As a result the effect of the $\alpha'$ corrections depends strongly on the mass and contributes to the calculation of the condensate. The overall better agreement of the lower temperature curve to the theoretical predictions is another signature that the observed deviations at large masses are due to $\alpha'$ corrections as opposed to lattice effects, although at sufficiently high masses ($|m^a| \lesssim 1/a$) lattice effect also become significant.

Note that this remarkable agreement (in the black hole phase) is obtained without any parameter fitting in contrast to the analogous studies of the BFSS matrix model~\cite{Hanada:2008ez}, where the authors performed a fit to estimate the $\alpha'$ corrections to the internal energy. We believe that it is the cancelation mechanism described above, which allows this highly non-trivial test of the gauge/gravity correspondence.  

The validity of our studies is justified by the lack of a serious sign problem in our lattice model. Similarly to the BFSS matrix model~\cite{Filev:2015hia}, one can show that only the real part of the Pfaffian contributes to the path integral and hence if the phase, $\theta$,  of the Pfaffian is in the range $-\pi/2<\theta<\pi/2$ there is no sign problem. Although there are configurations which violate this condition, numerical studies show that these are rare and the model does not suffer of a serous sign problem.
 
One may hope to improve the agreement in the Minkowski phase by going to lower temperatures. However, as usual for this system, at low temperature the model develops an instability due to flat directions, which requires larger size matrices. In addition, the condensate experiences significant fluctuations due to critical slowing down and the associated large aoutocorrelation times. Nevertheless, we plan to extend our numerical studies in this direction.


\section{Conclusion}

In this paper we performed a precision test of holographic flavour
dynamics. We focused on the study of a one-dimensional flavoured
Yang-Mills theory holographically dual to the D0/D4--brane
intersection, also known as the Berkooz--Douglas matrix model.  We
considered a lattice discretisation of the model which avoids fermion
doubling\footnote{Note that this was possible because the theory is
  one dimensional.}. Furthermore, super-renormalizability of the model
ensures that in the continuum limit, supersymmetry is broken only by
the effect of finite temperature, which enabled us to simulate it on a
computer.

Our results for the condensate versus bare mass curve show an
excellent agreement with holography in the regime of small bare masses
and at lower temperature this agreement extends to the whole range of
masses in the deconfined phase. We believe that this agreement can be
explained by a cancelation of the $\alpha'$ corrections to the
condensate for black hole embeddings (deconfined phase). This allows a
direct comparison between computer simulations and AdS/CFT predictions
at relatively high temperatures compared to similar studies of the
pure BFSS matrix model.

For Minkowski embeddings (confined phase) we observe significant
deviations from holography even for bare masses well bellow the lattice UV
cut-off, $1/a$. This disagreement is expected, since
for the temperatures that we study $\alpha'$ corrections to the free
energy are significant and, which is more important, vary
significantly with the bare mass parameter resulting in a significant
contribution to the fundamental condensate (unlike the deconfined
phase). An obvious way to improve the agreement in the confined phase
is to consider lower temperature when $\alpha'$ corrections become
less significant. Such studies are computationally very demanding due
to the large size matrices required to stabilise the model at low
temperature and the critical slowing down when approaching the gapless
zero temperature phase of the BFSS degrees of freedom. Alternatively we could 
attempt to estimate the $\alpha'$ corrections to the background along the lines of 
ref.~\cite{Kawahara:2007ib} (see also ref.~\cite{OtheraAlpha}). However, the main difficulty 
would be estimating the $\alpha'$ corrections to the DBI action, since not all such corrections 
are known in a curved background. Nevertheless we could attempt to estimate the mass and temperature 
dependance of the corrections and obtain the corresponding coefficients by fitting (in analogy to the studies of 
ref.~\cite{Kawahara:2007ib}.) We leave such studies for future work.

Another test of both our numerical approach and holography comes from
calculating the slope of the condensate curve, namely the
susceptibility $\partial^2F/\partial m_q^2$. This can be calculated
numerically by measuring the fluctuations of the condensate and the
expectation value of some appropriate operators. Our preliminary
studies for small bare masses showed satisfying agreement with the
slope of the condensate curve predicted by holography. We are
currently working on refining these studies.

Finally, in addition to the agreement to holography at low
temperature, we plan to verify our code by comparing to the
high-temperature expansion of the model~\cite{High-T-BD}. Our
preliminary results for the internal energy show excellent
agreement. We leave the more detailed and systematic study of other
observables for a future work.

{\bf Acknowledgements:} We wish to thank Yuhma Asano for useful comments and discussions. Part of the simulations were performed within the ICHEC ``Discovery" project ``dsphy003c".

\appendix
\section{Derivation of the holographic dictionary}
In this appendix we derive the second equation in (\ref{dictionary}). To obtain expression for the condensate as a function of the bare mass we use the definition: 
\begin{equation}
\langle {\cal O}_{m_q}\rangle \equiv \frac{\delta F}{\delta m_q}\ .
\end{equation}
The free energy of the theory can be obtained from the regularised Wick rotated on-shell action of the probe D4--brane. The approach that we take is to use an appropriate subtraction scheme\footnote{The same result can be obtained by introducing covariant counter term at the asymptotic boundary \cite{Benincasa:2009ze}, \cite{Karch:2005ms}.}. To this end we define new radial coordinate $\rho$ and a new field $L(\rho)$ via:
\begin{eqnarray}\label{L-rho}
\rho &=& \left(\frac{u^{7/2}+\sqrt{u^7-u_0^7}}{2}\right)^{2/7}\,\cos\theta\ , \\
L &=& \left(\frac{u^{7/2}+\sqrt{u^7-u_0^7}}{2}\right)^{2/7}\,\sin\theta\ . \nonumber
\end{eqnarray}
in these coordinates the DBI action (\ref{DBI-Wick}) can be written as:
\begin{eqnarray}
\label{DBI-Wick-L-rho}
S_{\rm DBI}^E &=&\frac{N_f\,\beta}{8\,\pi^2\,\alpha'^{5/2}\,g_s}\int\, d\rho\,\rho^3\,V\left(\rho,L(\rho)\right)\sqrt{1+L'(\rho)^2}\ .\nonumber \\
V\left(\rho,L(\rho)\right)&\equiv&\left(1-\frac{u_0^7}{4(\rho^2+L(\rho)^2)^{7/2}}\right)\,\left(1+\frac{u_0^7}{4(\rho^2+L(\rho)^2)^{7/2}}\right)^{1/7}\,
\end{eqnarray}
At large radial distances ($\rho \gg 1$) the solution for $L(\rho)$ has the expansion:
\begin{equation}
L(\rho) = u_0\,\tilde m+\frac{u_0^3\,\tilde c}{\rho^2}+\dots\ ,
\end{equation}
where the parameters $\tilde m$ and $\tilde c$ are the same as in equation (\ref{Expansion-sin(theta)}). The important property of these choice of coordinates is that if one introduces a UV cut-off $\rho_{\rm max}$ in the limit $\rho_{\rm max}\to \infty$ one gets:
\begin{equation}
S_{\rm DBI}^E =\frac{N_f\,\beta}{8\,\pi^2\,\alpha'^{5/2}\,g_s}\left(\frac{1}{4}\rho_{\rm max}^4+O(\rho_{\rm max}^0)\right)
\end{equation}
And hence the divergent term is independent on the parameters $\tilde m$ and $\tilde c$. Therefore, we can choose a simple subtraction scheme to regulate the action. A natural choice is to subtract the trivial embedding $L(\rho)\equiv 0$, which one can check is a solution to the equation of motion for $L(\rho)$. This results in the following expression for the fundamental free energy:
\begin{equation}
F = \frac{N_f}{8\,\pi^2\,\alpha'^{5/2}\,g_s}\left\{\int\limits_{\rho_0}^{\infty}d\rho\,\rho^3\,\left(V\left(\rho,L(\rho)\right)\sqrt{1+L'(\rho)^2}-1\right)+\frac{1}{4}\left(\rho_0^4-u_0^4\right)\right\}\ ,
\end{equation}
where $\rho_0 = \sqrt{{u_0^2}/2^{2/7}-L(\rho_0)^2}$ for black hole embeddings and vanishes $\rho_0=0$ for Minkowski embeddings (for $L(0) > {u_0^2}/2^{2/7}$). Using the relation $m_q = u_0\,\tilde m/(2\pi\alpha')$ for the condensate we obtain:
\begin{equation}
\langle {\cal O}_{m_q}\rangle \equiv {2\pi\alpha'}\frac{\delta F}{\delta L(\rho)}=\frac{N_f}{4\,\pi\,\alpha'^{3/2}\,g_s}\,\rho^3\,V\left(\rho,L(\rho)\right)\frac{L'(\rho)}{\sqrt{1+L'(\rho)^2}}\bigg|_{\rho_0}^{\infty}=-\frac{N_f\,u_0^3}{2\,\pi\,\alpha'^{3/2}\,g_s}\,\tilde c\ ,
\end{equation}
which is the expression for the condensate in equation (\ref{dictionary}).

\section{BFSS model. Discretisation}
\subsection{Bosonic action}
The bosonic part of the BFSS action is given by:
\begin{equation}
S_b = \frac{1}{g^2}\,\int_0^{\beta}dt\,{\rm tr}\left\{\frac{1}{2}({\cal D}_t {X^i})^2-\frac{1}{4}{\rm tr}[X^i,X^j]^2\right\}\ .
\end{equation}
Next we discretise time to $\Lambda$ sites $t_n = a n$, ($n =0,\dots,\Lambda-1$), where the lattice spacing is $a=\beta/\Lambda$ and the point $t_{\Lambda}=\Lambda a =\beta$ is identified with the point $0$. To discretise the covariant derivative ${\cal D}_t$ we define the transporters:
\begin{equation}\label{linkU}
U_{n,n+1} ={\cal P}\exp\left[i\int_{n a}^{(n+1)a}dt\,A(t)\right]\ ,
\end{equation}
where ${\cal P}$ denotes a path ordered product. Let us consider for a moment the pure derivative part of ${\cal D}_t$.  On the lattice we have:
\begin{equation}
 \partial_t X_n^i\rightarrow\frac{X_{n+1}^{i}-X_n^i}{a}\ .
\end{equation}
To make the above expression gauge covariant we have to transport back the field at $t_{n+1}$ to $t_n$. For the discrete version of the covariant derivative, we obtain:
\begin{equation}\label{cov_dev}
{\cal D}_t\rightarrow \frac{U_{n,n+1}X_{n+1}^iU_{n+1,n} -X_n^i}{a}\ ,
\end{equation}
where $U_{n+1,n} = U_{n,n+1}^{\dagger}$. Using equation (\ref{cov_dev}) for the discrete bosonic action we obtain:
\begin{equation}\label{Sb-discr}
S_b = N\sum_{n=0}^{\Lambda-1}{\rm tr}\left\{-\frac{1}{a}\,X_n^iU_{n,n+1}X^i_{n+1}U_{n,n+1}^{\dagger}+\frac{1}{a}\, (X_n^i)^2-\frac{a}{4}\,[X_n^i,X_n^j]^2\right\}\ ,
\end{equation}
where without loss of generality we have taken $g =\frac{1}{\sqrt {N}}$.\footnote{This can always be arranged by an appropriate resealing of the matrices and the time coordinate.} The action $S_b$ can be written in a much simpler form by using the ${\rm U}(n)^\Lambda$ gauge symmetry of the model. Indeed, at each lattice site we have a local ${\rm U}(N)$ symmetry. Using that symmetry we can perform the transformation:
\begin{eqnarray}\label{gauge-tr}
&&{X'}_0^i  =X_0^i\ ,\\ 
 &&{X'}_1^i = U_{0,1}\,X_1^i\,U_{0,1}^{\dagger}\ , \nonumber \\
 &&{X'}_2^i = (U_{0,1}U_{1,2})\,X_2^i\,(U_{0,1}U_{1,2})^{\dagger}\ ,\nonumber\\
 &&\dots\nonumber \\
&&{X'}_{\Lambda-1}^i = (U_{0,1}U_{1,2}\dots U_{\Lambda-2,\Lambda-1})\,X_{\Lambda-1}^i\,(U_{0,1}U_{1,2}\dots U_{\Lambda-2,\Lambda-1})^{\dagger}\nonumber \, 
\end{eqnarray}
introducing the notation ${\cal W} =(U_{0,1}U_{1,2}\dots U_{\Lambda-2,\Lambda-1}U_{\Lambda-1,0})$ for the bosonic action (\ref{Sb-discr}) we obtain:
\begin{equation}\label{Sb-discr-1}
S_b= -\frac{1}{a}N{\rm tr}\left\{\sum_{n = 0}^{\Lambda-2}{X'}_n^i{X'}^i_{n+1}+{X'}_{\Lambda-1}^i{\cal W}\,{X'}_0^i{\cal W}^{\dagger}\right\}+N\sum_{n=0}^{\Lambda-1}{\rm tr}\left\{\frac{1}{a}\, ({X'}_n^i)^2-\frac{a}{4}\,[{X'}_n^i,{X'}_n^j]^2\right\}\ .
\end{equation}
The unitary matrix ${\cal W}$ has the decomposition ${\cal W}=VD V^{\dagger} $, where $D ={\rm diag}\{e^{i\theta_1},\dots,e^{i\theta_N}\}$ is a diagonal unitary matrix and $V$ is a unitary. But the action (\ref{Sb-discr-1}) has the residual symmetry ${X'}_n^i \rightarrow V {X'}_n^i V^{\dagger} $, which we can use to diagonalise ${\cal W}$:
\begin{equation}\label{Sb-discr-D}
S_b= -\frac{1}{a}N{\rm tr}\left\{\sum_{n = 0}^{\Lambda-2}{X'}_n^i{X'}^i_{n+1}+{X'}_{\Lambda-1}^i{D}\,{X'}_0^i{D}^{\dagger}\right\}+N\sum_{n=0}^{\Lambda-1}{\rm tr}\left\{\frac{1}{a}\, ({X'}_n^i)^2-\frac{a}{4}\,[{X'}_n^i,{X'}_n^j]^2\right\}\ .
\end{equation}
We use this form of the action for coding on a computer.  

We could also use the additional symmetry ${X'}_n^i \rightarrow h_n {X'}_n^i h_n^{\dagger} $, where $h_n$ is a diagonal unitary matrix, to ``distribute'' the diagonal matrix $D$ among all of the hop terms. Indeed, defining the matrix $D_\Lambda ={\rm diag}\{e^{i\theta_1/\Lambda},\dots,e^{i\theta_N/\Lambda}\}$, which satisfies $(D_{\Lambda})^{\Lambda}=D$, one can verify that under the transformation: 
\begin{equation}\label{transformation}
{X'}_n^i = (V h_n ) \tilde X_n^i (V h_n)^{\dagger}  ~~~,{\rm where:}~~h_n = (D_{\Lambda})^n\ ,
\end{equation}
the action (\ref{Sb-discr-1}) transforms into:
\begin{equation}\label{Sb-discr-2}
S_b[\tilde X,D_{\Lambda}] = N\sum_{n=0}^{\Lambda-1}{\rm tr}\left\{-\frac{1}{a}\,\tilde X_n^iD_{\Lambda}\tilde X^i_{n+1}D_{\Lambda}^{\dagger}+\frac{1}{a}\, (\tilde X_n^i)^2-\frac{a}{4}\,[\tilde X_n^i,\tilde X_n^j]^2\right\}\ .
\end{equation}
Now let us discuss the measure of the transporter fields $U_{n,n+1}$. The measure can be written as:
\begin{equation}
\prod_{n =0}^{\Lambda-1} {\cal D}U_{n,n+1} = \prod_{n =1}^{\Lambda-1} {\cal D}U_{n,n+1} \, {\cal D}U_{0,1} = \prod_{n =1}^{\Lambda-1} {\cal D}U_{n,n+1} \, {\cal D}{\cal W}\ , 
\end{equation}
where we have used the that $U_{0,1} = {\cal W}\,(U_{1,2}\dots U_{\Lambda-2,\Lambda-1})^{\dagger}$ and the translational invariance of the measure. But the action (\ref{Sb-discr-2}) depends only on the matrix ${\cal W}$ (infact only on the eigenvalues of ${\cal W}$). Therefore the integration over the measure of the transporter fields results to:
\begin{eqnarray}
&&\int \prod_{n =0}^{\Lambda-1} {\cal D}U_{n,n+1} e^{-S_b[\tilde X,D_{\Lambda}]} =({Vol}\, {\rm U}(N))^{\Lambda-1}\int {\cal D W} e^{-S_b[\tilde X,D_{\Lambda}]}\propto \\
&\propto&\int\prod_{k =1}^Nd\theta_k\,\prod_{l >m}|e^{i\theta_l}-e^{i\theta_m}|^2\,e^{-S_b[\tilde X,D_\Lambda(\theta)]}  \propto \int\prod_{k =1}^Nd\theta_k\,e^{-S_b[\tilde X,D_\Lambda(\theta)]-S_{\rm FP}[\theta]} \nonumber\ ,
\end{eqnarray}
where $S_{\rm FP}[\theta]$ is given by:
\begin{equation}
S_{\rm FP} [\theta]= -\sum_{l\neq m}\ln\left|\sin \frac{\theta_l-\theta_m}{2}\right|\ .
\end{equation}
\subsection{Fermionic action}
The fermonic part of the action is:
\begin{equation}
S_f = \frac{1}{2g^2}\int d\tau \,{\rm tr}\left\{\psi^{\alpha} C_{9\,\alpha\beta}\,{\cal D}_\tau\psi^{\beta} -\psi^{\alpha} (C_{9}\gamma^i)_{\alpha\beta}\,[X^i,\psi^{\beta}]\right\}\ .
\end{equation}
We begin by splitting the fermions into two eight component fermions: $\psi=(\psi_1,\psi_2)$ and defining the forward and backward derivatives $D_\pm$:
\begin{eqnarray}
({\cal D}_-W)_n&=&(W_n-U_{n,n-1}W_{n-1}U_{n-1,n})/{a} \ , \nonumber \\
({\cal D}_+W)_n&=&(U_{n,n+1}W_{n+1}U_{n+1,n}-W_n)/{a}\ .
\end{eqnarray}
One can show that the discretised kinetic term then becomes:
\begin{eqnarray}
S_f^{\rm kin}&=&\frac{1}{2g^2}\int d\tau\,{\rm tr}\left(\psi^{\alpha} C_{9\,\alpha\beta}\,{\cal D}_\tau\psi^{\beta}\right) = \frac{a}{2g^2}\sum_{n=0}^{\Lambda-1}{\rm tr}\left\{\psi_{1,n}^T({\cal D}_-\psi_2)_n+\psi_{2,n}^T({\cal D}_+\psi_1)_n    \right\} =\\
&&=\frac{1}{g^2}{\rm tr}\left\{-\sum_{n=0}^{\Lambda-1}\psi^T_{2,n}\psi_{1,n}+\sum_{n=0}^{\Lambda-2}\psi^T_{2,n}U_{n,n+1}\psi_{1,n+1}U_{n+1,n} \pm \psi^T_{2,\Lambda-1}U_{\Lambda-1,0}\psi_{1,0}U_{0,\Lambda-1}\right\}\,\,\nonumber\ ,
\end{eqnarray}
where the plus/minus sign in the last term corresponds to periodic/anti-periodic boundary conditions for the fermions.\footnote{Namely the conditions $\psi_{-1}=\pm\psi_{\Lambda-1}$ and  $\psi_{\Lambda}=\pm\psi_{0}$. }
Using the gauge from the previous subsection when the hollonomy is concentrated on a singe link we can write $S_f^{\rm kin}$ as:
\begin{equation}
S_f^{\rm kin}=\frac{1}{g^2}{\rm tr}\left\{-\sum_{n=0}^{\Lambda-1}\psi^T_{2,n}\psi_{1,n}+\sum_{n=0}^{\Lambda-2}\psi^T_{2,n}\psi_{1,n+1}\pm \psi^T_{2,\Lambda-1}D\,\psi_{1,0}\,D^{\dagger}\right\}\ .
\end{equation}
Since all fields transform in the adjoint of $SU(N)$ instead of dealing with matrices we can use the corresponding real components: $X^a={\rm tr} (T^aX)$ and $\psi^a = {\rm tr} (T^a\psi)$, where $T^a$ are the standard basis of $SU(N)$ normalised as ${\rm tr}\, T^aT^b =\delta^{ab}$. $S_f^{\rm kin}$ can then be written as:
\begin{eqnarray}
S_f^{\rm kin} &=& \frac{1}{g^2}\sum_{a,b =0}^{N^2-1}\sum_{m,n =0}^{\Lambda-1}\sum_{\alpha = 1}^8\psi^{\alpha+8}_{m\,,a}K^{ab}_{mn}\psi^{\alpha}_{n\,,b}\ , \\
K_{mn}^{ab}&=&(\delta_{m+1,n} -\delta_{m,n})\delta^{ab}\pm\delta_{m,\Lambda-1}\delta_{n,0}\,d^{ab}\\
d^{ab} &=&{\rm tr}\left(T^a\,D\,T^b\,D^{\dagger}\right)\ .
\end{eqnarray}
where the plus/minus sign corresponds to periodic/ant-periodic boundary conditions. The kinetic term can also be written as:
\begin{eqnarray}
S_f^{\rm kin} &=&\sum_{a,b =0}^{N^2-1}\sum_{m,n =0}^{\Lambda-1}\sum_{\alpha,\beta = 1}^{16}\psi^{\alpha}_{m\,,a}\,{\cal M}^{\rm kin}_{mn,\alpha\beta,ab}\,\psi^{\beta}_{n\,,b}\\
{\cal M}^{\rm kin}_{mn,\alpha\beta,ab} &=&\frac{1}{2g^2}\left(\begin{array}{cc} 0_8 &  -K_{nm}^{ba}\\K_{mn}^{ab} & 0_8 \end{array}\right)_{\alpha\beta}\ .
\end{eqnarray} 
Discretising the potential part of the action is straightforward. One obtains:
\begin{eqnarray}
S_f^{\rm pot} &=&\sum_{a,b =0}^{N^2-1}\sum_{m,n =0}^{\Lambda-1}\sum_{\alpha,\beta = 1}^{16}\psi^{\alpha}_{m\,,a}\,{\cal M}^{\rm pot}_{mn,\alpha\beta,ab}\,\psi^{\beta}_{n\,,b}\\
{\cal M}^{\rm pot}_{mn,\alpha\beta,ab} &=&\frac{1}{2g^2}\,a\,if^{abc}\,(C_{9}\gamma^i)_{\alpha\beta}\,X^{c,i}_n\ .
\end{eqnarray}
Finally, defining:
\begin{eqnarray}\label{M-psi}
{\cal M}_{mn,\alpha\beta,ab} &=&\frac{1}{2g^2}\left(\begin{array}{cc} 0_8 &  -K_{nm}^{ba}\\K_{mn}^{ab} & 0_8 \end{array}\right)_{\alpha\beta}+\frac{1}{2g^2}\,a\,if^{abc}\,(C_{9}\gamma^i)_{\alpha\beta}\,X^{c,i}_n\ .
\end{eqnarray}
We can write:
\begin{equation}\label{S-psi}
S_f = \psi^T{\cal M}\psi\ .
\end{equation}


\begin{thebibliography}{99}

\bibitem{Maldacena:1997re} 
  J.~M.~Maldacena,
  \atmp{2}{1998}{231}
  \arXivid{hep-th/9711200}.
  
\bibitem{Karch:2002sh} 
  A.~Karch and E.~Katz,
  \jhep{0206}{2002}{043}
  \arXivid{hep-th/0205236}.
  

\bibitem{Babington:2003vm} 
  J.~Babington, J.~Erdmenger, N.~J.~Evans, Z.~Guralnik and I.~Kirsch,
  \prd{69}{2004}{066007 }
  \arXivid{hep-th/0306018}.
  
 
\bibitem{Hoyos:2006gb} 
  C.~Hoyos-Badajoz, K.~Landsteiner and S.~Montero,
  \jhep{0704}{2007}{031}
  \arXivid{hep-th/0612169}.

\bibitem{Mateos:2007vn} 
D.~Mateos, R.~C.~Myers and R.~M.~Thomson,
  \prl{97}{2006}{091601}
  \arXivid{hep-th/0605046}.\\
  D.~Mateos, R.~C.~Myers and R.~M.~Thomson,
  \jhep{0705}{2007}{067}
  \arXivid{hep-th/0701132}.
   
\bibitem{Mateos:2007vc}
  D.~Mateos, S.~Matsuura, R.~C.~Myers and R.~M.~Thomson,
  \jhep{0711}{2007}{085}
  \arXivid{0709.1225} [hep-th].
  
\bibitem{EM-field} 
  V.~G.~Filev, C.~V.~Johnson, R.~C.~Rashkov and K.~S.~Viswanathan,
  \jhep{0710}{2007}{019}
  \arXivid{hep-th/0701001}.\\
  T.~Albash, V.~G.~Filev, C.~V.~Johnson and A.~Kundu,
  \jhep{0808}{2008}{092}
  \arXivid{0709.1554} [hep-th].\\
  J.~Erdmenger, R.~Meyer and J.~P.~Shock,
  \jhep{0712}{2007}{091}
  \arXivid{0709.1551} [hep-th].
  
\bibitem{Erdmenger:2007ja} 
  J.~Erdmenger, M.~Kaminski and F.~Rust,
  \prd{77}{2008}{046005 }
  \arXivid{0710.0334} [hep-th].
  
\bibitem{Evans:2008nf} 
  N.~Evans and E.~Threlfall,
  \prd{79}{2009}{066008}
  \arXivid{0812.3273} [hep-th].
  
\bibitem{O'Bannon:2008bz} 
  A.~O'Bannon,
  \jhep{0901}{2009}{074}
  \arXivid{0811.0198} [hep-th].
  
\bibitem{Karch:2015kfa}
  A.~Karch, B.~Robinson and C.~F.~Uhlemann,
  \prl{115}{2015}{261601}
  \arXivid{1509.00013} [hep-th].
  
\bibitem{Catterall:2005fd} 
  S.~Catterall,
  \jhep{0506}{2005}{027}
  \arXivid{hep-lat/0503036}.
  
\bibitem{Kaplan:2005ta} 
  D.~B.~Kaplan and M.~Unsal,
  \jhep{0509}{2005}{042}
  \arXivid{hep-lat/0503039}.
  
\bibitem{Anagnostopoulos:2007fw} 
  K.~N.~Anagnostopoulos, M.~Hanada, J.~Nishimura and S.~Takeuchi,
  \prl{100}{2008}{021601}
  \arXivid{0707.4454} [hep-th].

    \bibitem{Catterall:2008yz} 
  S.~Catterall and T.~Wiseman,
  \prd{78}{2008}{041502}
  \arXivid{0803.4273} [hep-th].
  
\bibitem{Hanada:2008ez} 
  M.~Hanada, Y.~Hyakutake, J.~Nishimura and S.~Takeuchi,
  \prl{102}{2009}{191602}
  \arXivid{0811.3102} hep-th].
  
\bibitem{Catterall:2009xn}
  S.~Catterall and T.~Wiseman,
  \jhep{1004}{2010}{077}
  \arXivid{0909.4947} [hep-th].
  
\bibitem{Hanada:2013rga}
  M.~Hanada, Y.~Hyakutake, G.~Ishiki and J.~Nishimura,
  Science {\bf 344} (2014) 882
  \arXivid{1311.5607} hep-th].

\bibitem{Kadoh:2015mka} 
  D.~Kadoh and S.~Kamata,
 \arXivid{1503.08499} [hep-lat].

\bibitem{Filev:2015hia} 
  V.~G.~Filev and D.~O'Connor,
 \arXivid{1506.01366} [hep-th].
  
\bibitem{Joseph:2015xwa} 
  A.~Joseph,
  \ijmpa{30}{2015}{1530054}
  \arXivid{1509.01440} [hep-th].
  
\bibitem{Itzhaki:1998dd} 
  N.~Itzhaki, J.~M.~Maldacena, J.~Sonnenschein and S.~Yankielowicz,
  \prd{58}{1998}{046004 }
  \arXivid{hep-th/9802042}.
  
  \bibitem{First-order}
  I. Kirsch, 
  \forp{52}{2004}{727} \\
  T.~Albash, V.~G.~Filev, C.~V.~Johnson and A.~Kundu,
  \prd{77}{2008}{066004}
  \arXivid{hep-th/0605088]}.

\bibitem{Berkooz:1996is} 
  M.~Berkooz and M.~R.~Douglas,
  \plb{395}{1997}{196}
\arXivid{hep-th/9610236}.
  
\bibitem{VanRaamsdonk:2001cg} 
  M.~Van Raamsdonk,
  \jhep{0202}{2002}{001}
 \arXivid{hep-th/0112081]}.
  
 \bibitem{Davies}
 R.~Davies,
 http://people.maths.ox.ac.uk/daviesr/resources/notes/spinors.pdf

\bibitem{Clark:2004cp} 
  M.~A.~Clark, A.~D.~Kennedy and Z.~Sroczynski,
  \npps{140}{2005}{835}
  \arXivid{hep-lat/0409133}.
 
\bibitem{Benincasa:2009ze} 
  P.~Benincasa,
  \arXivid{0903.4356} [hep-th].
  
\bibitem{Karch:2005ms} 
  A.~Karch, A.~O'Bannon and K.~Skenderis,
  \jhep{0604}{2006}{015}
  \arXivid{hep-th/0512125}
  
\bibitem{Kawahara:2007ib} 
  N.~Kawahara, J.~Nishimura and S.~Takeuchi,
 \jhep{0712}{2007}{103}
  \arXivid{0710.2188} [hep-th].
  
  \bibitem{OtheraAlpha}
    S.~S.~Gubser, I.~R.~Klebanov and A.~A.~Tseytlin,
    \npb {534}{1998}{202},
    \hepth{9805156}.

    A.~Buchel, J.~T.~Liu, A.~O.~Starinets,
    \npb {707}{2005}{56-68},
    \hepth{0406264}.

  A.~Buchel,
  \npb {803}{2008}{166},
  \arXivid{0805.2683} [hep-th].

    B.~Hassanain and M.~Schvellinger,
    \jhep {1201}{2012}{114},
    \arXivid{1108.6306}.

    B.~Hassanain and M.~Schvellinger,
    \prd {85}{2012}{086007},
    \arXivid{1110.0526} [hep-th].

  D.~L.~Yang and B.~M\"uller,
  \arXivid{1507.04232} [hep-th].

  S.~Grozdanov and A.~O.~Starinets,
  \jhep {1503}{2015}{007},
  \arXivid{1412.5685} [hep-th].

  S.~Waeber, A.~Sch\"afer, A.~Vuorinen and L.~G.~Yaffe,
  \jhep{1511}{2015}{087}
  \arXivid{1509.02983} [hep-th].

  \bibitem{High-T-BD}
  Y.~Asano, V.~Filev, S.~Kovacik, D.~O'Connor ({\it to appear})

\end{thebibliography}
\end{document}